\documentclass{JHEP3}
\usepackage{amsmath,amssymb}
\usepackage{multirow}

\newcommand{\CB}{{\cal B}}
\newcommand{\CO}{{\cal O}}
\newcommand{\CU}{{\cal U}}
\newcommand{\CL}{{\cal L}}
\newcommand{\CN}{{\cal N}}
\newcommand{\bCN}{\bar{\CN}}

\newcommand{\Nd}{N^\dag}
\newcommand{\del}{{\partial}}
\newcommand{\nn}{\nonumber \\}
\newcommand{\NR}{{\rm NR}}
\newcommand{\bmsigma}{\mbox{\boldmath{$\sigma$}}}
\newcommand{\bmalpha}{\mbox{\boldmath{$\alpha$}}}
\newcommand{\bmp}{\mbox{\boldmath{$p$}}}
\newcommand{\textoverline}[1]{$\overline{\mbox{#1}}$}

\allowdisplaybreaks

\title{Four-nucleon contact interactions from holographic QCD}

\author{
Youngman Kim,$^{1,2}$ Deokhyun Yi$^{1,2}$ and Piljin Yi$^{3}$\\

$^1$Asia Pacific Center for Theoretical Physics,\\
$\,$Pohang, Gyeongbuk 790-784, Korea

$^2$Department of Physics, Pohang University of Science and Technology,\\
$\,$Pohang, Gyeongbuk 790-784, Korea

$^3$School of Physics, Korea Institute for Advanced Study,\\
$\,$Seoul 130-722, Korea\\

E-mail: \email{ykim@apctp.org}, \email{dada@postech.ac.kr}, \email{piljin@kias.re.kr}}

\preprint{KIAS-P11067}

\abstract{We calculate the low energy constants of four-nucleon interactions in an effective chiral Lagrangian in holographic QCD.
We start with a D4-D8 model to obtain meson-nucleon interactions and then integrate out massive mesons to obtain
the four-nucleon interactions in 4D.
We end up with two low energy constants at the leading order and seven of them at the next leading order,
which is consistent with the effective chiral Lagrangian.
The values of the low energy constants are evaluated with the first five Kaluza-Klein resonances.}


\begin{document}

\section{Introduction}

The four-nucleon interaction in the effective chiral Lagrangian was first introduced by Weinberg \cite{Weinberg:1990rz}
to describe the short-range part of the nuclear force, see \cite{Machleidt:2011zz} and references therein
for a recent review on nuclear forces from low-energy QCD via chiral effective field theory.
Like pion and pion-nucleon interactions in chiral Lagrangian, the four-Fermi operators
are accompanied by unknown coupling constants, called low energy constants (LECs). These constants are calculable from Quantum Chromodynamics (QCD), in principle.
However, in reality the LECs are determined by a fit to some experimental data or through some model-dependent calculations.
Various LECs in the meson sector of a low-energy effective chiral lagrangian involving pseudoscalar fields only were
determined with the resonance saturation hypothesis~\cite{Ecker:1988te};
the assumption that dominant contributions to the LECs come from the dynamics of the low-lying resonances.
In~\cite{Epelbaum:2001fm}, the coupling constants of four-nucleon interactions are calculated  with the resonance saturation assumption.

The four-nucleon terms are also important to understand the bulk nuclear matter property using the chiral Lagrangian.
For instance, it was shown that the repulsive vector mean field in the Walecka model can be identified by the Four-Fermi interaction
in chiral Lagrangian~\cite{WChPT}. This indicates that the chiral Lagrangian with the four-nucleon terms
could be satisfactorily describing the bulk property of nuclear matter as the Walecka model~\cite{Serot:1984ey} does in the mean field approximation.
The role of the four-Fermi interaction in establishing a bridge between the chiral quark-meson coupling model and chiral Lagrangian within the mean field approximation
was discussed in~\cite{Saito:2010zw}.
The effect of the four-Fermi interaction on pion and kaon condensation was delved into in~\cite{mesonC}.

In this work we evaluate the LECs in a D4-D8 model with explicit bulk baryon fields \cite{Hong:2007kx, Hong:2007ay, Kim:2009sr}.
We start from the 4D meson-baryon Lagrangian obtained in~\cite{Hong:2007kx, Hong:2007ay, Kim:2009sr}. We integrate out massive mesons
to obtain 4D relativistic four-nucleon contact interactions.  Then, we take a non-relativistic limit to arrive at the LECs in the
effective chiral Lagrangian.  In the sense that integrated-out massive mesons determine the value of the LECs, our approach is similar to
\cite{Epelbaum:2001fm} based on the resonance saturation assumption.
We consider up to $Q^2$ order, where $Q$ is a typical momentum of a system at hand.
At the leading order ($Q^0$) we have two LECs and seven of them at $Q^2$ order.  We evaluate the nine LECs  with contributions
from the first five Kaluza-Klein (KK) modes.

\section{A D4/D8 holographic QCD and meson-nucleon couplings}

Holography \cite{AdSCFT} is a conjectural property of string theory,
whereby a strongly coupled large $N_c$ gauge theory can be recast into
a weakly coupled closed string theory. In practice, one starts
with a D-brane configuration that contains large $N_c $ gauge
theory as its low energy sector, study how closed string theory
view the configuration, and identify the degrees of freedom
in the latter as the gauge-invariant objects (color-singlets)
of the former. As such, we end up with an effective description
involving objects like glueballs, mesons, and baryons in case
of QCD.

A prototypical and probably the best candidate so far is
D4/D8 model. In this configuration, one starts with $N_c$
D4-branes, but now compactified on a thermal circle $S^1$
of radius $1/M_{KK}$,
where one requires anti-periodic boundary conditions
on all fermions along the circle, breaking all supersymmetry.
Four dimensional remnant is precisely flavorless $U(N_c)$
QCD at lowest energy level. The holography views this as a
ten-dimensional supergravity theory in the background of
\cite{Gibbons:1987ps}
\begin{equation}
ds^2=\left(\frac{U}{R}\right)^{3/2}\left(\eta_{\mu\nu}dx^{\mu}dx^{\nu}+f(U)d\tau^2\right)
+\left(\frac{R}{U}\right)^{3/2}\left(\frac{dU^2}{f(U)}+U^2d\Omega_4^2\right) \;,
\end{equation}
with $f(U) \equiv 1- U_{KK}^3/U^3$. $x^\mu$ and $\tau$
are the directions along which the D4-brane is extended.
The most central quantity is $M_{KK}=2\pi/\delta\tau=
3U_{KK}^{1/2}/2R^{3/2}$, which is not only the scale associated
with the circle $\tau$ but also both the curvature
scale at $U=U_{KK}$ and the scale associated with the
internal $S^4$ manifold. $R$ is more directly related
to the underlying string construction
as $R^3=\pi g_s N_c l_s^3$ with the string
coupling $g_s$ and string length scale $l_s$.
$U$ is the holographic direction and bounded from below
by the condition $U\geq U_{KK}$.

Taking large $N_c$ limit naturally decouples $U(1)$ part
of $U(N_c)$ in the usual $1/N_c$ expansion sense, so
we have $SU(N_c)$ theory at hand. Analyzing the type IIA
supergravity theory in this background, near $U\sim U_{KK}$,
is supposed to be equivalent to studying $SU(N_c)$ QCD without quarks
at very low energy. However, one must understand how
mesons and baryons are realized in this context,
which will be the subjects of next two subsections.

\subsection{Mesons}\label{sec:Mesons}

For mesons and baryons, one must add quarks to the above. It turns out that
the right thing to do is to introduce D8 and \textoverline{D8} pairs of
branes. With $N_f$ pairs, QCD with $N_f$ massless quarks
emerges, although from the holographic side one only sees
mesons. On the latter viewpoint, the D8's and the \textoverline{D8}'s
are located at $\tau=0$ and $\tau=\pi \delta\tau$, so in
fact identified at $U=U_{KK}$. This gluing represents the
chiral symmetry breaking in a geometric manner, since what
would have been $U(N_f)\times U(N_f)$ symmetry becomes
$U(N_f)$ instead.
This is the celebrated Sakai-Sugimoto model~\cite{sakai-sugimoto}.

In the end, the meson spectrum and interactions are all realized
as the lowest-lying flavor gauge theory on D8 branes.
Instead of going through how this construction works, let us
summarize the meson sector in terms of a five-dimensional $U(N_f)$
flavor gauge theory on the worldvolume (modulo $S^4$ which
we are ignoring) of the now connected D8 and \textoverline{D8},
\begin{eqnarray}\label{fiveM}
&&-\frac14\;\int d^4xdw\;
\; \frac{1}{e(w)^2}\;{\rm tr} {\cal F}^2
+\frac{N_c}{24\pi^2}\int_{4+1}\omega_{5}({\cal A})
\end{eqnarray}
with the Chern-Simons 5-form, $\omega_5({\cal A})$, and
\begin{equation}
\frac{1}{e(w)^2}
=\frac{\lambda N_c}{108\pi^3} \,u(w)\,M_{KK}\: .
\end{equation}
The holographic radial coordinate $U$ is exchanged in favor of
$w$, related to $u=U/U_{KK}$ as,
\begin{equation}
 \frac23 \,w M_{KK}=\pm \int_{1}^u{dy}/\sqrt{y^3-1}\ .
\end{equation}

This five-dimensional action contains an infinite
tower of spin 1 mesons and the pseudo-scalar field
$U$ as
\begin{equation}
{\cal A}_\mu(x;w)= i\,[U^{-1/2},\partial_\mu U^{1/2}]/2 +i\{U^{-1/2},\partial_\mu U^{1/2}\}\psi_0(w)
+\sum_n v_\mu^{(n)}(x)\psi_{(n)}(w)\:
\end{equation}
with eigenfunctions $\psi_{(n)}(w)$ along $w$ directions.
The gauge choice here is ${\cal A}_w=0$, which although not
entirely sensible suffices for the purpose here. The zero-mode
part $U(x)$ can be thought of as the open Wilson line, $
U(x)=exp({i\int_w {\cal A}})$, in generic gauge but here
was gauge-transformed into ${\cal A}_\mu$.

The quadratic terms in (\ref{fiveM}) produces two type of terms
in four dimensions
\begin{eqnarray}\label{ex}
&&\int d^4x \;\left(\frac{f_\pi^2}{4}{\rm tr} \left(U^{-1}\partial_\mu
U\right)^2 +\frac{1}{32 e^2_{Skyrme}} {\rm tr} \left[ U^{-1}\partial_\mu U,
U^{-1} \partial_\nu U \right]^2\right) \nonumber\\ \nonumber \\
&+&\int d^4x \sum_{n=1}^\infty{\rm tr}\; \left\{\frac{1}{2} (dv^{(n)})_{\mu\nu}
(dv^{(n)})^{\mu\nu}+m_{(n)}^2 v_\mu^{(n)} v^{(n)\mu}\right\}+\cdots\:,
\end{eqnarray}
with $ f_\pi^2=(g_{YM}^2 N_c) N_c M_{KK}^2/54\pi^4$ and
$1/e^2_{Skyrme}\simeq  {61 (g_{YM}^2 N_c)N_c}/54\pi^7$.
For real QCD, $M_{KK}$ would be roughly $M_{KK}\sim m_{\CN}\sim 0.94 \mathrm{GeV}$,
while $f_\pi \sim 93 \mathrm{MeV}$, and this requires $(g_{YM}^2N_c)N_c \sim 50$.
For $N_c=3$, this gives $\lambda=g_{YM}^2N_c\simeq 17$.
Suppressed here are interaction terms among these mesons,
cubic, quartic, and also Wess-Zumino-Witten term which
comes from $w_5({\cal A})$. This theory of holographic
mesons has been investigated much in the literature.
We emphasize that apart from the two input parameters $\lambda$
and $M_{KK}$, no other tunable parameter exists. All the masses
and all the couplings are fixed unambiguously via this
Kaluza-Klein process from five dimensions to four dimensions.
This remarkable aspect will persist to baryon sector in next
subsection.

For later purpose, we wish to identify four-dimensional mesons
more clearly. We assume $SU(2)$ isospin symmetry and
separate out the iso-singlet and the iso-triplet sector as
\begin{equation}
v_\mu^{(2k-1)}=
\omega^{(k)}_\mu\left(\begin{array}{cc}1/2 &0 \\ 0&1/2\end{array}\right)+\frac12\,\rho_\mu^{(k)a}\tau^a\:,
\qquad
v_\mu^{(2k)}=
f^{(k)}_\mu\left(\begin{array}{cc}1/2 &0 \\ 0&1/2\end{array}\right)+\frac12\, a_\mu^{(k)a}\tau^a\:,
\end{equation}
for vectors and axial-vectors, respectively.
We will sometimes use the notation
$\rho_\mu=\rho_\mu^a\tau^a$ and $a_\mu=a_\mu^a\tau^a$ also.
Even/odd nature of $\psi_{(n)}(w)$ translates to the
usual parity of the corresponding mesons, so vectors,
$\rho$'s and $\omega$'s, and axial vectors, $a_1$'s and $f_1$'s,
alternates in the infinite tower of massive spin 1 mesons.
On the other hand, Goldstone bosons associated with chiral symmetry
breaking reside in the open Wilson line as
\begin{equation}
U=exp(\pi i (\eta'+\pi^a\tau_a)/f_\pi)\ .
\end{equation}
The $U(1)$ part $\eta'$ picks up a mass term,
\begin{equation}\label{eta mass}
m_{\eta'}^2=\frac{1}{27\pi^2}\frac{N_f}{N_c}\lambda^2M_{KK}^2
\end{equation}
via a holographic version of axial anomaly. We refer readers
to Ref.~\cite{sakai-sugimoto} for derivation of this additional
contribution.

\subsection{Holographic baryons}

Baryons are identified as the soliton of the above flavor gauge theory,
characterized by the first Pontryagin number,
\begin{equation}
p_1({\cal F})\equiv \frac{1}{8\pi^2}\int_{x^{1,2,3},w}{\rm tr}\, {\cal F} \wedge {\cal F}=1 \,.
\end{equation}
For simplicity, let us consider $N_f=2$. Quantizing the soliton
to produce spin 1/2 baryons, and representing them via a local
field ${\cal B}$, baryon dynamics can be added to the flavor gauge theory
of meson sector as \cite{Hong:2007kx, Hong:2007ay}
\begin{eqnarray}\label{fiveB}
&&+\int d^4 x \,dw\left[-i\bar{\cal B}\gamma^m D_m {\cal B}
-i m_{\cal B}(w)\bar{\cal B}{\cal B} +\frac{g(w)^2\rho_{baryon}^2}{e^2(w)}\bar{\cal B}\gamma^{mn}F_{mn}{\cal B} \right] \
\end{eqnarray}
where
\begin{equation}\label{size}
\qquad m_{\cal B}(w)= \frac{4\pi^2}{e(w)^2},\qquad
\rho_{baryon}^2\simeq
\frac{({2\cdot 3^7\cdot\pi^2/5})^{1/2}}{M_{KK}^2\lambda } \ ,
\end{equation}
and the function $g_5(w)$ is known only at $w=0$, $g_5(0)=2\pi^2/3$,
which suffices for what follows in the large $\lambda$ and the large $ N_c$ limits.
We will not repeat how this action is derived from
D4/D8 holographic QCD, but note again that there is no
free parameter.

Note that the coupling between mesons and baryons are
achieved via two interaction terms. The first is embedded into
the covariant derivative,
\begin{equation}
D_m\equiv \partial_m-i(N_c {\cal A}_m^{U(1)}+{A}_m) \ ,
\end{equation}
for which the flavor gauge field  ${\cal A}_m$ of (\ref{fiveM})
is decomposed  as ${\cal A}_m^{U(1)}+A_m$ with
traceless $2\times 2$ $A_m$. The second is through
a direct coupling to the $SU(N_f=2)$  field strength, ${F}$,
which can be traced to the fact that the soliton underlying the
baryon carried the second Pontryagin number, $p_1(F)=1$, whose
configuration is self-dual magnetic fields in the four
spatial directions.

\subsection{Nucleon-meson dynamics and cubic couplings}

To obtain meson-nucleon interactions
we expand
\begin{equation}
{\cal B}(x^\mu,w)=\CN_{+}(x^\mu)f_{+}(w)+\CN_{-}(x^\mu)f_-(w)\, ,
\end{equation}
where $\gamma^5 \CN_{\pm}=\pm \CN_{\pm}$. The profile functions $f_\pm(w)$ satisfy
\begin{eqnarray}
\partial_w f_+(w)+m_{\cal B}(w) f_+(w) &=& m_{\cal N} f_-(w)\:,\nonumber\\
-\partial_w f_-(w)+m_{\cal B}(w) f_-(w) &=& m_{\cal N} f_+(w)\:
\label{f_pm}
\end{eqnarray}
where $m_{\cal N}$ is the nucleon mass in 4D.
The 4D Dirac field for the nucleon is given by
\begin{equation}
{\cal N}= \CN_+ + \CN_- \:.
\end{equation}
The eigenfunctions $f_\pm(w)$ are normalized as
\begin{equation}
\int_{-w_{max}}^{w_{max}} dw\,\left|f_+(w)\right|^2 =
\int_{-w_{max}}^{w_{max}} dw\,\left|f_-(w)\right|^2 =1\:,
\end{equation}
and the eigenvalue $m_{\CN}$ is the mass of the nucleon mode $\CN(x)$.
From (\ref{f_pm}), we get the second-order equations for $f_\pm(w)$
\begin{eqnarray}
\left[-\partial_w^2-\partial_w m_{\CB}(w)+m_{\CB}(w)^2\right] f_+(w)=m_{\CN}^2 f_+(w)\:,\nonumber\\
\left[-\partial_w^2+\partial_w m_{\CB}(w)+m_{\CB}(w)^2\right] f_-(w)=m_{\CN}^2 f_-(w)\:.
\label{eigeneq}
\end{eqnarray}
There is a 1-1 mapping of eigenmodes with $f_-(w)=\pm f_+(-w)$,
where the sign choice is related to the sign choice of $m_\CN$.
Due to the asymmetry under $w\rightarrow -w$, $f_+(w)$ tends to shift
to the positive $w$ side and the opposite happens for $f_-(w)$.
In this work we will take the convention $f_-(w)=f_+(-w)$.

Inserting this into the action (\ref{fiveB}), we find the following structure
of the four-dimensional nucleon action \cite{Kim:2009sr}
\begin{eqnarray}\label{eq:4d}
&&\int d^4x\;{\cal L}_4 = \int d^4x\left(-i\bar {\cal N}
\gamma^\mu\partial_\mu {\cal N}-im_{\cal N}\bar {\cal N}{\cal N}+ {\cal
L}_{\rm vector} +{\cal L}_{\rm axial}\right)\:,
\end{eqnarray}
where the couplings to mesons are
\begin{eqnarray} \label{coupling triplet}
 {\cal L}_{\rm vector}&=&
 -\sum_{k \ge 1}\frac{g_{V}^{(k)triplet}}{2} \bar {\cal N} \gamma^\mu  \rho_\mu^{(k)}{\cal N}
-\sum_{k \ge 1}\frac{N_cg_{V}^{(k)singlet}}{2} \bar {\cal N} \gamma^\mu  w_\mu^{(k)}{\cal N} \nn
&&+\sum_{k \ge 1}\frac{g_{dV}^{(k)triplet}}{2} \bar {\cal N} \gamma^{\mu\nu}  \partial_\mu \rho_\nu^{(k)}{\cal N} +\cdots\:,
\end{eqnarray}
and
\begin{eqnarray}\label{coupling singlet}
{\cal L}_{\rm axial}&=&\frac{ g_A^{triplet}}{2f_\pi}\bar {\cal N}  \gamma^\mu\gamma^5 \partial_\mu\pi {\cal N}
+\frac{ N_c g_A^{singlet}}{2f_\pi}\bar {\cal N}  \gamma^\mu\gamma^5 \partial_\mu\eta' {\cal N}  \nn \nn
&& -\sum_{k\ge 1}\frac{ g_A^{(k)}}{2} \bar {\cal N} \gamma^\mu\gamma^5
a_\mu^{(k)triplet} {\cal N} -\sum_{k\ge 1}\frac{ N_cg_A^{(k)singlet}}{2} \bar {\cal N} \gamma^\mu\gamma^5
f_\mu^{(k)} {\cal N}+\cdots
\end{eqnarray}
with $\pi=\pi^a\tau^a$. The ellipses denote quartic or higher terms.
A notable feature here
is that derivative couplings to spin 1 mesons exist only for $\rho$ mesons.
All others vanish as we will see presently.
The pseudoscalar coupling can be alternatively written as
\begin{equation}\label{coupling pi eta}
-\left(\frac{g_A^{triplet}}{2f_\pi}\times 2m_{\cal N}\right)\bar {\cal N} \gamma^5
\pi{\cal N}
- \left(\frac{g_A^{singlet} N_c}{2f_\pi}\times 2m_{\cal N}\right)\bar {\cal N} \gamma^5 \eta'{\cal N} \,
\end{equation}
using the on-shell condition of nucleons, which define $g_{\pi \CN\CN}$ and $ g_{\eta' \CN\CN}$.
Note that we are considering two flavor case
$N_f=2$;  $\eta'$ denotes the trace part of the
pseudo-scalar, regardless of the number of flavors.

Let us take a closer look at 4D cubic couplings.
Here we summarize the results from Refs.~\cite{Hong:2007kx,Hong:2007ay,Kim:2009sr}
following the notation in Ref.~\cite{Kim:2009sr}.
The 5D Lagrangian (\ref{fiveB}) generate 4D couplings via the minimal coupling
\begin{equation}
-\int dw \;\bar{\cal B}\gamma^m  (N_c {\cal A}_m^{U(1)}+{A}_m) {\cal B}\, ,
\end{equation}
and the derivative coupling
\begin{equation}
\int dw \;g_5(w)\frac{\rho_{baryon}^2}{e^2(w)}\bar{\cal B}\gamma^{mn}F_{mn}{\cal B},
\end{equation}
with $g_5(0)=2\pi^2/3$ \cite{Hong:2007ay}, upon mode-expanding the flavor gauge field
and retaining the  lowest-lying mode of $\CB$. For instance, the latter generates
two types of terms as
\begin{eqnarray}
\gamma^{\mu\nu}F_{\mu\nu}(x,w)
&=&2\sum_{n}\psi_{(n)}(w)\gamma^{\mu\nu}\partial_\mu \left[v_\nu^{(n)}(x)\biggl\vert_{\rm iso-triplet}\right]\:,\nn
\gamma^{5\mu}F_{5\mu}(x,w)
&=&-2\sum_{n}\left(\partial_w\psi_{(n)}(w)\right)\gamma^\mu\gamma^5\left[v_\mu^{(n)}\biggl\vert_{\rm iso-triplet}\right]\, ,
\end{eqnarray}
from which it is already clear that iso-singlet vectors $w$ and $f_1$ cannot have
derivative couplings to the nucleon in this approximation.
Integrating over $w$ will then give cubic couplings as overlap integrals involving
one $\psi$ and two $f_1$'s.

It is convenient to define three set of numbers
$A_n$, $B_n$ and $C_n$ as
\begin{align}\label{eq:An Bn Cn}
A_n &= \int_{-w_{max}}^{w_{max}} dw\,\left|f_+(w)\right|^2 \psi_{(n)}(w)\:, \nn
B_n &= \int_{-w_{max}}^{w_{max}}
 dw \left(g_5(w)\frac{\rho_{baryon}^2}{e^2(w)}\right)
\, f_-^*(w)f_+(w) \psi_{(n)}(w)\:, \nn
C_n &= \int_{-w_{max}}^{w_{max}}
 dw \left(g_5(w)\frac{\rho_{baryon}^2}{e^2(w)}\right)
\,\left|f_+(w)\right|^2 \partial_w \psi_{(n)}(w)\:
\end{align}
from which all cubic couplings to mesons are constructed.
$B$'s are in responsible for the derivative couplings.
Choosing the phase of nucleon eigenfunction as $f_+(w)=-f_-(-w)$,
and noting that $\psi_{(n)}(w)$ is even/odd function
when $n$ is even/odd, respectively, we see that $B_{2k}=0$.
This leads to the result that $a_1$ mesons have no derivative
coupling to nucleons.

Sometimes additional $\gamma^5$ is generated in terms originating
from vector-like 5D coupling, because
$\psi_{(2k)}$ and $\partial_w\psi_{(2k-1)}$ are odd functions of $w$
and $\gamma^5 \CN_{\pm }=\pm \CN_{\pm }$.
Taking all of these into account, one finds
\begin{eqnarray}
g_V^{(k)triplet}&=& A_{2k-1}+2C_{2k-1} \ , \nn
g_A^{(k)triplet}&=& 2C_{2k}+A_{2k}\ ,\nn
g_{dV}^{(k)triplet}&=& 2B_{2k-1}\ ,
\end{eqnarray}
and
\begin{eqnarray}
g_V^{(k)singlet}&=& A_{2k-1}\ , \nn
g_A^{(k)singlet}&=& A_{2k}\ .
\end{eqnarray}
Note that the vertices involving isospin singlet vector and axial-vector mesons are
constructed only by minimal coupling terms.
Cubic couplings to pseudo-scalars are determined similarly as
\begin{eqnarray}
g_A^{triplet}&=& 4C_0+2A_0 \ , \nn
g_A^{singlet}&=& 2A_0 \ .
\end{eqnarray}

\section{Four-nucleon contact interactions in chiral Lagrangian}

In this section we summarize the four-nucleon contact interactions in an effective chiral Lagrangian
which was first introduced by Weinberg \cite{Weinberg:1990rz}
to describe the short-range part of the nuclear force. Like pion and pion-nucleon interactions in chiral Lagrangian, the four-Fermi operators
are accompanied by LECs whose value are undetermined. These constants are calculable from QCD, in principle.
However, in practice the LECs are determined by a fit to some experimental data or through some model-dependent calculations.
In the next section, we will calculate the LECs using the meson-baryon vertex derived from the D4/D8/\textoverline{D8} model.

\subsection{Structures and LECs}

In the conventional one-boson exchange (OBE) model of the nucleon-nucleon ($NN$) force   long
range part is dominated by one-pion exchange, intermediate attraction is mostly given by a scalar meson,  and  short range interaction is
described in terms of the vector meson exchange. Note that the scalar does not necessarily be the chiral partner of the pion, and its effect can
be described by two-pion exchanges. In a modern approach for the $NN$ force based on a chiral effective Lagrangian,
(multi-) pion effects together with contact nucleon interactions replace the OBE picture.
We focus on the  four-nucleon contact interactions that can be expressed as  a sum of local operators
with increasing number of derivatives, or expansion in powers of a small momentum scale $Q$.

\renewcommand{\arraystretch}{1.2}
\begin{table}[!ht]
\centering
\parbox{.5\textwidth}{
{\small
\begin{tabular}{c|c|c}
\hline
\hline
$O_S^{}$  & $(\Nd N)(\Nd N)$ & leading\\
$O_T^{}$  & $(\Nd \bmsigma N)\cdot (\Nd\bmsigma N)$ & ($Q^0$)\\
\hline
$O_1^{}$  & $(\Nd \nabla N)^2 +{\rm h.c.}$  & \multirow{3}*{$\sigma \times 0$} \\
$O_2^{}$  & $(\Nd \nabla N )\cdot ( \nabla \Nd  N) $\\
$O_3^{}$  &  $(\Nd N) ( \Nd \nabla^2 N)+{\rm h.c.}$  \\
\hline
$O_4^{}$  & $i \,( \Nd \nabla N) \cdot (\nabla \Nd \times \bmsigma N )+ {\rm h.c.}$ & \multirow{3}*{$\sigma\times 1$}\\
$O_5^{}$ & $i \, (\Nd N)(\nabla \Nd \cdot \bmsigma \times \nabla N)$ \\
$O_6^{}$ & $i \, (\Nd \bmsigma N) \cdot (\nabla \Nd \times \nabla N)$\\
\hline
$O_7^{}$  & $( \Nd \bmsigma \cdot \nabla N) (\Nd \bmsigma\cdot \nabla N) +{\rm h.c.}$ & \multirow{8}*{$\sigma\times 2$}\\
$O_8^{}$  & $(\Nd \sigma^i \del_j N)(\Nd \sigma^j \del_i N) + {\rm h.c.}$  \\
$O_9^{}$  &   $(\Nd \sigma^i \del_j N)(\Nd \sigma^i \del_j N) + {\rm h.c.}$ \\
$O_{10}^{}$  & $(\Nd \bmsigma \cdot \nabla N)(\nabla \Nd \cdot\bmsigma N)$ \\
$O_{11}^{}$  & $(\Nd \sigma^i \del_j N)(\del_i \Nd \sigma^j N)$ \\
$O_{12}^{}$  &  $(\Nd \sigma^i \del_j N)(\del_j \Nd \sigma^i N)$\\
$O_{13}^{}$  &  $ (\Nd \sigma^i N)(\del_j \Nd \sigma^j \del_i N) +{\rm h.c.}$\\
$O_{14}^{}$  & $2(\Nd \sigma^i N)(\del_j \Nd \sigma^i \del_j N)$ \\
\hline
\hline
\end{tabular}}
\caption{Isosinglet operators}
\label{tb:OPlist}
}
\hfill
\parbox{.46\textwidth}{
{\small
\begin{tabular}{c|c}
\hline
\hline
$O_S^{\tau}$  & $(\Nd \tau^a N)(\Nd \tau^a N)$ \\
$O_T^{\tau}$  & $(\Nd \tau^a \bmsigma N)\cdot (\Nd \tau^a \bmsigma N)$ \\
\hline
$O_1^{\tau}$  & $(\Nd \tau^a \nabla N)^2 +{\rm h.c.}$ \\
$O_2^{\tau}$  & $(\Nd \tau^a \nabla N )\cdot ( \nabla \Nd \tau^a N) $ \\
$O_3^{\tau}$  &  $(\Nd \tau^a N) ( \Nd \tau^a \nabla^2 N)+{\rm h.c.}$ \\
\hline
$O_4^{\tau}$  & $i \,( \Nd \tau^a \nabla N) \cdot (\nabla \Nd \tau^a  \times \bmsigma N )+ {\rm h.c.}$\\
$O_5^{\tau}$ & $i \, (\Nd \tau^a N)  (\nabla \Nd \cdot \tau^a  \bmsigma \times \nabla N)$\\
$O_6^{\tau}$ & $i \, (\Nd \tau^a \bmsigma N) \cdot (\nabla \Nd \tau^a \times \nabla N)$\\
\hline
$O_7^{\tau}$  & $( \Nd \tau^a \bmsigma \cdot \nabla N) (\Nd \tau^a \bmsigma\cdot \nabla N) +{\rm h.c.}$ \\
$O_8^{\tau}$  & $(\Nd \tau^a \sigma^i \del_j N)(\Nd \tau^a \sigma^j \del_i N) + {\rm h.c.}$  \\
$O_9^{\tau}$  &   $(\Nd \tau^a \sigma^i \del_j N)(\Nd \tau^a \sigma^i \del_j N) + {\rm h.c.}$ \\
$O_{10}^{\tau}$  & $(\Nd \tau^a \bmsigma \cdot \nabla N)(\nabla \Nd \tau^a \cdot\bmsigma N)$ \\
$O_{11}^{\tau}$  & $(\Nd \tau^a \sigma^i \del_j N)(\del_i \Nd \tau^a \sigma^j N)$ \\
$O_{12}^{\tau}$  &  $(\Nd \tau^a \sigma^i \del_j N)(\del_j \Nd \tau^a \sigma^i N)$\\
$O_{13}^{\tau}$  &  $ (\Nd \tau^a \sigma^i N)(\del_j \Nd \tau^a \sigma^j \del_i N) +{\rm h.c.}$\\
$O_{14}^{\tau}$  & $2(\Nd \tau^a \sigma^i N)(\del_j \Nd \tau^a \sigma^i \del_j N)$ \\
\hline
\hline
\end{tabular}}
\caption{Isotriplet operators}
\label{tb:OPlist triplet}
}
\end{table}
\renewcommand{\arraystretch}{1.0}

At the leading order (LO) ($Q^0$), the expansion is the four (non-relativistic) nucleon  interactions
with no derivatives \cite{Weinberg:1990rz}
\begin{equation}
\CL^{(0)}=-\frac{1}{2}C_S (N^\dagger N)(N^\dagger N)
-\frac{1}{2}C_T (N^\dagger \bmsigma N)\cdot (N^\dagger\bmsigma N),
\end{equation}
where $N$ is the two component nucleon field and
$C_S$ and $C_T$ denote the low energy constants (LECs).
At  $Q^2$ order, the contact Lagrangian can be written as \cite{Ordonez:1995rz}
\begin{equation}
\CL^{(2)}=-\sum_{i=1}^{14}C_i'O_i,
\end{equation}
where $C_i'$ are LECs and $O_i$ are 14 operators listed in Table~\ref{tb:OPlist}.
Then, the four-point contact Lagrangian up to $Q^2$ is
\begin{equation}\label{eq:Lagrangian}
\CL = -\frac{1}{2}C_S O_S -\frac{1}{2}C_T O_T-\sum_{i=1}^{14}C_i'O_i \,.
\end{equation}
The 14 operators in Table~\ref{tb:OPlist} are the isosinglet $(I=0)$ operators
and the isotriplet $(I=1)$ operators are in Table~\ref{tb:OPlist triplet}.
Note that only 12 out of these 14 operators are independent since
\begin{equation}\label{op_independence}
O_7-O_8=2O_{11}-2O_{10},\quad O_4+O_5=O_6\, .
\end{equation}
Using the Fierz identity (Appendix~\ref{sec:Fierz}),
we can rewrite each isotriplet operator in terms of isosinglet operators as
{\small
\begin{align}\label{operators triplet}
O_S^{\tau} & = O_S \,, \nn
O_T^{\tau} & = 3 O_S - 2 O_T \,, \nn
O_1^{\tau} & = O_1 \,, \nn
O_2^{\tau} & = -O_1 -3O_2 -O_3 \,, \nn
O_3^{\tau} & = O_3 \,, \nn
O_4^{\tau} & = -3 O_4 \,, \nn
O_5^{\tau} & = O_4 - O_5 - O_{10} + O_{11} \,, \nn
O_6^{\tau} & = -O_4 - O_6 - O_{10} + O_{11} \,, \nn
O_7^{\tau} & = O_1 + 2O_5 + 2O_6 + O_8 - O_9 \,, \nn
O_8^{\tau} & = O_1 - 2O_5 - 2O_6 + O_7 - O_9 \,, \nn
O_9^{\tau} & = 3O_1 - 2O_9 \,, \nn
O_{10}^{\tau} & = -\frac{1}{2}(O_1 + 2O_2 + O_3) - O_5 - O_6 - O_{10} + O_{13} - \frac{1}{2}O_{14} \,, \nn
O_{11}^{\tau} & = -\frac{1}{2}(O_1 + 2O_2 + O_3) + O_5 + O_6 - O_{11} + O_{13} - \frac{1}{2}O_{14} \,, \nn
O_{12}^{\tau} & = -\frac{3}{2}(O_1 + 2O_2 + O_3) - O_{12} - \frac{1}{2}O_{14} \,, \nn
O_{13}^{\tau} & = 2O_2 + 2O_{10} + 2O_{11} - 2O_{12} - O_{13} \,, \nn
O_{14}^{\tau} & = 6O_2 - 2O_{12} - O_{14} \,.
\end{align}
}We confirm the relation
\begin{equation}
O_7^{\tau} - O_8^{\tau} = 2O_{11}^{\tau} - 2O_{10}^{\tau},\quad
O_4^{\tau} + O_5^{\tau} = O_6^{\tau} \,
\end{equation}
and again only 12 out of these 14 operators are independent
for the isotriplet sector.

\subsection{Non-relativistic limit}\label{sec:Non-relativistic}

The relativistic fermion field $\CN$ can be reduced to the
two-component spinor $N$ via nonrelativistic expansion.
For this, it is convenient to use a Dirac basis
\begin{equation*}
\gamma^0=\left(\begin{array}{rr} -i & 0 \\ 0 & i\end{array}\right),\quad
\gamma^i=\left(\begin{array}{cc} 0 & -\sigma_i \\ -\sigma_i & 0\end{array}\right),\quad
\gamma^5=\left(\begin{array}{rr} 0 & -i \\ i & 0\end{array}\right)
\end{equation*}
whereby we have an expansion
\begin{equation}\label{eq:CN N}
  \CN(x) = \left( \begin{array}{c}
N(x)+\frac{1}{8m_\CN^2}\nabla^2N(x)\\
\frac{1}{2m_\CN}\bmsigma\cdot\nabla N(x)
 \end{array} \right)
 + {\cal O}(Q^3).
\end{equation}
In the leading $Q^0$ order, this becomes
\begin{equation}\label{eq:CN N 0}
  \CN(x) = \left( \begin{array}{c}
  N(x) \\
 0
 \end{array} \right).
\end{equation}
 Since we are to
compute contact terms up to dimension eight operators, we  retain
higher order correction in the upper component. Details of
this expansion is reviewed in Appendix~\ref{sec:NonrelSpinor}
for the sake of completeness.

One consequence of building quartic contact terms from
reduction of relativistic interaction vertices is that not
all of $O$'s in Table~\ref{tb:OPlist} appears independently. The underlying
Lorentz symmetry constrains the contact terms such that only
nine (2+7) linearly independent combinations appear up to dimension eight.
These are \cite{Girlanda:2010ya}
\begin{align}\label{eq:A7op}
{\cal A}_S&=O_S + \frac{1}{4m_\CN^2}(O_1 + O_3 + O_5 + O_6)\,,\nn
{\cal A}_T&=O_T - \frac{1}{4m_\CN^2}( O_5 +  O_6 - O_7 + O_8 +  2\, O_{12} +  O_{14})\,,\nn
{\cal A}_1&=O_1 + 2\, O_2\,,\quad
{\cal A}_2=2\, O_2 + O_3\,,\quad
{\cal A}_3=O_9 + 2\, O_{12}\,,\nn
{\cal A}_4&=O_9 + O_{14}\,,\quad
{\cal A}_5= O_5 - O_6\,,\nn
{\cal A}_6&=O_7 + 2\, O_{10}\,,\quad
{\cal A}_7=O_7 + O_8  + 2\, O_{13}
\end{align}
which consist of two leading operators (of $Q^0$ order) with higher order corrections
and seven subleading ones (of $Q^2$ order).
The effective Lagrangian corresponding to (\ref{eq:Lagrangian}) can be written as
\begin{eqnarray}\label{eq:EFT Lagrangian with As}
{\cal L} &=& -\frac{1}{2} C_S {\cal A}_S- \frac{1}{2} C_T {\cal A}_T
  - \frac{1}{2} C_1 {\cal A}_1 + \frac{1}{8} C_2 {\cal A}_2 - \frac{1}{2} C_3 {\cal A}_3 \nonumber \\
&&- \frac{1}{8} C_4 {\cal A}_4- \frac{1}{4} C_5 {\cal A}_5 - \frac{1}{2} C_6 {\cal A}_6
-\frac{1}{16} C_7 {\cal A}_7\, .
\end{eqnarray}
This is the same contact Lagrangian given in \cite{Ordonez:1995rz}
in terms of $O_{i=1\ldots 14}$ operators in Table~\ref{tb:OPlist}.
The representation of seven independent coupling constants
$C_{1,\ldots , 7}$ of $Q^2$ order Lagrangian is also in agreement
with the result of  \cite{Epelbaum:2001fm, Epelbaum:2000kv}
by using the reparametrization invariance \cite{Luke:1992cs}.
For isotriplet sector, we have   {\small
\begin{align}\label{eq:A7op triplet}
{\cal A}_S^{\tau}&=O_S^{\tau} + \frac{1}{4m_\CN^2}(O_1^{\tau} + O_3^{\tau} + O_5^{\tau} + O_6^{\tau}) \nn
 &=O_S + \frac{1}{4m_\CN^2}(O_1 + O_3 - O_5 - O_6 +O_7 - O_8) \nn
 &={\cal A}_S + \frac{1}{4m_\CN^2}(- 2O_5 - 2O_6 + O_7 - O_8) \,,\nn
{\cal A}_T^{\tau}&=O_T^{\tau} - \frac{1}{4m_\CN^2}( O_5^{\tau} +  O_6^{\tau} - O_7^{\tau} + O_8^{\tau} +  2\, O_{12}^{\tau} +  O_{14}^{\tau})\nn
 &=3O_S -2O_T - \frac{1}{4m_\CN^2}( -3O_1 - 3O_3 -5O_5 -5O_6 +2O_7 -2O_8 -4O_{12} -2O_{14} )\nn
 &=3{\cal A}_S -2{\cal A}_T \,,\nn
{\cal A}_1^{\tau}&=O_1^{\tau} + 2\, O_2^{\tau}
 = - O_1 -6O_2 -2O_3
 = - {\cal A}_1 - 2{\cal A}_2 \,,\nn
{\cal A}_2^{\tau}&=2\, O_2^{\tau} + O_3^{\tau}
 = -2O_1 -6O_2 -O3
 = - 2{\cal A}_1 - {\cal A}_2 \,,\nn
{\cal A}_3^{\tau}&=O_9^{\tau} + 2\, O_{12}^{\tau}
 = -6O_2 -3O_3 -2O_9 -2O_{12} -O_{14}
 = - 3{\cal A}_2 - {\cal A}_3 - {\cal A}_4 \,,\nn
{\cal A}_4^{\tau}&=O_9^{\tau} + O_{14}^{\tau}
 = 3O_1 + 6O_2 -2O_9 -2O_{12} - O_{14}
 = 3{\cal A}_1 - {\cal A}_3 - {\cal A}_4 \,,\nn
{\cal A}_5^{\tau}&= O_5^{\tau} - O_6^{\tau}
 = -3O_5 + 3O_6
 = -3{\cal A}_5 \,,\nn
{\cal A}_6^{\tau}&=O_7^{\tau} + 2\, O_{10}^{\tau}
 = -2O_2 - O_3 +O_8 -O_9 -2O_{10} + 2O_{13} - O_{14} \nn
 &= -{\cal A}_2 - {\cal A}_4 - {\cal A}_6 + {\cal A}_7 \,,\nn
{\cal A}_7^{\tau}&=O_7^{\tau} + O_8^{\tau}  + 2\, O_{13}^{\tau}
 = 2O_1 + 4O_2  + 3O_7 -O_8 -2O_9 + 8O_{10} -4O_{12} -2O_{13} \nn
 &= 2{\cal A}_1 - 2{\cal A}_3 + 4{\cal A}_6 - {\cal A}_7 \,.
\end{align} \par}
In this work, we will construct the contact term by integrating out massive mesons in holographic QCD.
We  then  obtain relativistic
quartic operators  after the integrating-out, such as $\bCN\CN\bCN\CN$, $\bCN\gamma^\mu\CN\bCN\gamma_\mu\CN$,
$\bCN\gamma^\mu\gamma^5\CN\bCN\gamma_\mu\gamma^5\CN$, and so on, and
expand them into quartic operators involving $N$'s. Here we list the
result of such expansion for all relativistic quartic operators we
will encounter in the next section:
{\small
\begin{align}\label{eq:operators}
\bCN\gamma^{\mu}\CN \bCN\gamma_\mu\CN
  \rightarrow&\: -O_S + \frac{1}{4 m_\CN^2} \left( 4O_2 + 2O_5 - 4O_6 - O_7 + O_9 - 2O_{10} + 2O_{12} \right) \nn
  &= -{\cal A}_S + \frac{1}{4 m_\CN^2} \left( {\cal A}_1 + {\cal A}_2 + {\cal A}_3 + 3{\cal A}_5 - {\cal A}_6 \right) \,, \nn
\bCN \gamma^{\mu}\CN\del^2\left(\bCN \gamma_\mu \CN\right)
  \rightarrow&\: O_1 + 2O_2
  = {\cal A}_1 \,, \nn
\bCN\gamma^\mu\gamma^5\CN \bCN\gamma_\mu\gamma^5\CN
  \rightarrow&\: O_T + \frac{1}{4 m_\CN^2}\left(- 2O_6 + O_7 - O_9 - 2O_{10} - 2O_{12} + 2O_{13} - 2O_{14}\right) \nn
  &= {\cal A}_T + \frac{1}{4 m_\CN^2}\left(- {\cal A}_4 + {\cal A}_5 - {\cal A}_6 + {\cal A}_7 \right) \,, \nn
\bCN \gamma^\mu\gamma^5\CN\del^2\left(\bCN \gamma_\mu\gamma^5 \CN\right)
  \rightarrow&\: O_9 + 2O_{12}
  = {\cal A}_3 \,, \nn
\del_\mu\left(\bCN\gamma^\mu\gamma^5\CN\right)\del_\nu\left(\bCN\gamma^\nu\gamma^5\CN\right)
  \rightarrow&\: O_7 + 2O_{10}
  = {\cal A}_6 \,, \nn
\bCN\gamma_{\mu}\CN \partial_\nu\left(\bCN\gamma^{\nu\mu}\CN\right)
  \rightarrow&\: \frac{1}{2m_\CN^{}}\left( O_1 + 2O_2 + 2O_5 - 2O_6 - O_7 + O_9 - 2O_{10} + 2O_{12} \right) \nn
  &= \frac{1}{2m_\CN^{}}\left( {\cal A}_1 + {\cal A}_3 - 2{\cal A}_5 - {\cal A}_6 \right) \,, \nn
\partial_\nu\left(\bCN{\gamma^\nu}_\mu\CN\right)\partial_\lambda\left(\bCN\gamma^{\lambda\mu}\CN\right)
  \rightarrow&\: -O_7 + O_9 - 2O_{10} + 2O_{12}
  = {\cal A}_3 - {\cal A}_6\, .
\end{align}
}Here we used the equation of motion $i \gamma^\mu\partial_\mu \CN+im_\CN\CN=0$
to eliminate time derivatives and performed partial integrations.
Similarly we arrive at, for isotriplet sectors,
{\small
\begin{align}\label{eq:operators triplet}
\bCN\gamma^{\mu}\tau^a\CN \bCN\gamma_\mu\tau^a\CN
  \rightarrow &\:
  -{\cal A}_S^\tau + \frac{1}{4 m_\CN^2} \left( {\cal A}_1^\tau + {\cal A}_2^\tau + {\cal A}_3^\tau + 3{\cal A}_5^\tau - {\cal A}_6^\tau \right) \nn
  &= -{\cal A}_S + \frac{1}{4 m_\CN^2} \left( -3{\cal A}_1 - 5{\cal A}_2 - {\cal A}_3 -9{\cal A}_5 +{\cal A}_6 -{\cal A}_7 \right) \,, \nn
\bCN \gamma^{\mu}\tau^a\CN\del^2\left(\bCN \gamma_\mu \tau^a\CN\right)
  \rightarrow&\:
  {\cal A}_1^\tau
  = -{\cal A}_1 -2{\cal A}_2 \,, \nn
\bCN\gamma^\mu\gamma^5\tau^a\CN \bCN\gamma_\mu\gamma^5\tau^a\CN
  \rightarrow&\:
  {\cal A}_T^\tau + \frac{1}{4 m_\CN^2}\left(- {\cal A}_4^\tau + {\cal A}_5^\tau - {\cal A}_6^\tau + {\cal A}_7^\tau \right) \nn
  &= 3{\cal A}_S - 2{\cal A}_T \nn
  & \quad + \frac{1}{4 m_\CN^2} \left( -{\cal A}_1 + {\cal A}_2 - {\cal A}_3 + 2{\cal A}_4 -3{\cal A}_5 +5{\cal A}_6 -2{\cal A}_7 \right) \,, \nn
\bCN \gamma^\mu\gamma^5\tau^a\CN\del^2\left(\bCN \gamma_\mu\gamma^5 \tau^a\CN\right)
  \rightarrow&\:
  {\cal A}_3^\tau
  = -3{\cal A}_2 -{\cal A}_3 -{\cal A}_4 \,, \nn
\del_\mu\left(\bCN\gamma^\mu\gamma^5\tau^a\CN\right)\del_\nu\left(\bCN\gamma^\nu\gamma^5\tau^a\CN\right)
  \rightarrow&\:
  {\cal A}_6^\tau
  = -{\cal A}_2 - {\cal A}_4 - {\cal A}_6 + {\cal A}_7 \,, \nn
\bCN\gamma_{\mu}\tau^a\CN \partial_\nu\left(\bCN\gamma^{\nu\mu}\tau^a\CN\right)
  \rightarrow&\:
  \frac{1}{2m_\CN^{}}\left( {\cal A}_1^\tau + {\cal A}_3^\tau - 2{\cal A}_5^\tau - {\cal A}_6^\tau \right) \nn
  &= \frac{1}{2m_\CN^{}}\left( -{\cal A}_1 -4{\cal A}_2 -{\cal A}_3 +6{\cal A}_5 +{\cal A}_6 -{\cal A}_7 \right) \,, \nn
\partial_\nu\left(\bCN{\gamma^\nu}_\mu\tau^a\CN\right)\partial_\lambda\left(\bCN\gamma^{\lambda\mu}\tau^a\CN\right)
  \rightarrow&\:
  {\cal A}_3^\tau - {\cal A}_6^\tau
  = -2{\cal A}_2 -{\cal A}_3 +{\cal A}_6 -{\cal A}_7  \, ,
\end{align}
}where we have ignored the higher order ($m_\CN^{-2}$) corrections of ${\cal A}_S^\tau$ from (\ref{eq:A7op triplet}).

\section{Four-nucleon contact interactions in holographic QCD}

We now calculate the LECs in (\ref{eq:EFT Lagrangian with As}) in holographic QCD.
We start from the 4D meson-baryon Lagrangian in (\ref{eq:4d}) derived from the Sakai-Sugimoto model with explicit bulk baryon field
~\cite{Hong:2007ay, Kim:2009sr}.
Then we integrate out massive mesons to obtain the values of the LECs.
In the sense that integrated-out massive mesons determine the value of the LECs, our approach is similar to
\cite{Epelbaum:2001fm} based on the resonance saturation assumption.

\subsection{Isospin singlet mesons}
We first consider the contributions from isospin singlet mesons to the LECs.

\subsubsection*{$\omega$ meson}

For the isosinglet vector meson $\omega$, the interaction is shown in (\ref{coupling triplet}).
The relativistic effective Lagrangian for the baryon field $\CN$ with $\omega$ meson is
\begin{align}\label{L omega CN}
&\, -\sum_{k\ge 1}
\left(
\frac{1}{4}(dw^{(k)})_{\mu\nu}(dw^{(k)})^{\mu\nu}+\frac{1}{2}m_{\omega^{(k)}}^2\omega^{(k)}_\mu\omega^{(k)\mu}\right)\nn
&\, -\sum_{k\geq1}
\left(\frac{N_c g_{V}^{(k)singlet}}{2}\right)\bar {\CN} \gamma^\mu \omega^{(k)}_\mu \CN\, ,
\end{align}
where $dw^{(k)}_{\mu\nu}=\del_\mu\omega^{(k)}_\nu-\del_\nu\omega^{(k)}_\mu$.
Solving the equation of motion for $\omega^{(k)\mu}$,
\begin{align}
\left(-m_{\omega^{(k)}}^2+\del^2\right)\omega^{(k)}_\mu-\del_\mu\del_\lambda \omega^{(k)\lambda}
=\left(\frac{N_c g_{V}^{(k)singlet}}{2}\right)\bCN \gamma_\mu \CN \,,
\end{align}
we find $\del_\mu\omega^{(k)\mu} = 0$ from the equations of motions of $\CN$ and $\bCN$.
Then by performing a derivative expansion, we obtain
\begin{eqnarray}\label{eq:omega}
\omega_\mu^{(k)}
&=& -\frac{1}{m_{\omega^{(k)}}^2}\left(\frac{N_c g_{V}^{(k)singlet}}{2}\right)\bCN\gamma_\mu\CN \nn
&&{}\, -\frac{1}{m_{\omega^{(k)}}^4}\left(\frac{N_c g_{V}^{(k)singlet}}{2}\right)\del^2\left(\bCN\gamma_\mu\CN\right)
\,+\CO(Q^3)\,.
\end{eqnarray}
By substitution this into (\ref{L omega CN}),
we arrive at the contact interaction due to the $\omega$ meson exchange
\begin{eqnarray}
\CL_{\omega}&=&
\sum_{k\geq1}
\frac{1}{2m_{\omega^{(k)}}^2}
\left(\frac{N_c g_{V}^{(k)singlet}}{2}\right)^2
\bCN \gamma^{\mu} \CN \bCN \gamma_\mu\CN\nn
&&+\sum_{k\geq1}
\frac{1}{2m_{\omega^{(k)}}^4}
\left(\frac{N_c g_{V}^{(k)singlet}}{2}\right)^2
\bCN \gamma^{\mu}\CN\del^2\left(\bCN \gamma_\mu \CN\right)
\,+\CO(Q^3)\, .
\end{eqnarray}
If we use the non-relativistic reduction (\ref{eq:operators}),
this becomes
\begin{eqnarray}\label{N N omega}
\CL_{\omega}&\rightarrow&
-\sum_{k\geq1}
\frac{1}{2m_{\omega^{(k)}}^2}
\left(\frac{N_c g_{V}^{(k)singlet}}{2}\right)^2 {\cal A}_S \nn
&&+\frac{1}{4 m_\CN^2}\sum_{k\geq1}\frac{1}{2m_{\omega^{(k)}}^2}
\left(\frac{N_c g_{V}^{(k)singlet}}{2}\right)^2
\left( {\cal A}_1 + {\cal A}_2 + {\cal A}_3 + 3{\cal A}_5 - {\cal A}_6 \right)
\nn
&&+\sum_{k\geq1}
\frac{1}{2m_{\omega^{(k)}}^4}
\left(\frac{N_c g_{V}^{(k)singlet}}{2}\right)^2
{\cal A}_1\,.
\end{eqnarray}

\subsubsection*{$f_1$ meson}
For the isospin singlet axial-vector meson $f_1$,
the interaction is
\begin{eqnarray}
-\sum_{k\geq1}
\left(\frac{N_cg_{A}^{(k)singlet}}{2}\right)
\bCN\gamma^{\mu}\gamma^5 f_\mu^{(k)}\CN\,.
\end{eqnarray}
Similar to the $\omega$ meson case, we integrate out $f_1$ meson using the
equation of motion and
\begin{align}
\del_\mu f^{(k)\mu}=
-\frac{1}{m_{f^{(k)}}^2}\left(\frac{N_cg_{A}^{(k)singlet}}{2}\right)
\del_\mu \left( \bCN \gamma^\mu\gamma^5 \CN \right)
\end{align}
to obtain
\begin{eqnarray}
\CL_{f_1}&=&
\sum_{k\geq1}
\frac{1}{2m_{f^{(k)}}^2}
\left(\frac{N_c g_{A}^{(k)singlet}}{2}\right)^2
\bCN\gamma^\mu\gamma^5\CN \bCN\gamma_\mu\gamma^5\CN \nn
&&+\sum_{k\geq1}
\frac{1}{2m_{f^{(k)}}^4}
\left(\frac{N_c g_{A}^{(k)singlet}}{2}\right)^2
\bCN \gamma^\mu\gamma^5\CN\del^2\left(\bCN \gamma_\mu\gamma^5 \CN\right) \nn
&&+\sum_{k\geq1}
\frac{1}{2m_{f^{(k)}}^4}
\left(\frac{N_c g_{A}^{(k)singlet}}{2}\right)^2
\del_\mu\left(\bCN \gamma^\mu\gamma^5 \CN\right)\del_\nu\left(\bCN \gamma^\nu\gamma^5 \CN\right)
\,+\CO(Q^3)\,.\,
\end{eqnarray}
Again from (\ref{eq:operators}), we get the non-relativistic reduced form
\begin{eqnarray}\label{N N f}
\CL_{f_1}&\rightarrow&
\sum_{k\geq1}
\frac{1}{2m_{f^{(k)}}^2}
\left(\frac{N_c g_{{\cal A}}^{(k)singlet}}{2}\right)^2 {\cal A}_T \nn
&&+\frac{1}{4 m_\CN^2}\sum_{k\geq1}\frac{1}{2m_{f^{(k)}}^2}
\left(\frac{N_c g_{{\cal A}}^{(k)singlet}}{2}\right)^2
\left( -{\cal A}_4 + {\cal A}_5 - {\cal A}_6 + {\cal A}_7\right) \nn
&&+\sum_{k\geq1}\frac{1}{2m_{f^{(k)}}^4}
\left(\frac{N_c g_{{\cal A}}^{(k)singlet}}{2}\right)^2
\left( {\cal A}_3 + {\cal A}_6 \right)\,.
\end{eqnarray}

\subsubsection*{$\eta'$ meson}
For the pseudo-scalar meson $\eta'$ case,
\begin{eqnarray}
+\left(\frac{N_cg_{A}^{(k)singlet}}{2f_\pi}\right)\bar{\CN}\gamma^\mu\gamma^5\del_\mu\eta'\CN \ ,
\label{L eta CN}
\end{eqnarray}
the equations of motion of $\eta'$ gives
\begin{eqnarray}
\eta'
=-\frac{1}{m_{\eta'}^2}
\left(\frac{N_cg_A^{singlet}}{2f_\pi}\right)
\del_\mu(\bCN\gamma^\mu\gamma^5\CN) \,+ \CO(Q^3) \,.
\end{eqnarray}
Then, the interaction can be reduced into the
following four-point interactions
\begin{align}
\CL_{\eta'}
=\frac{1}{2m_{\eta'}^2}
\left(\frac{N_cg_{A}^{singlet}}{2f_\pi}\right)^2
\del_\mu\left(\bCN\gamma^\mu\gamma^5\CN\right)
\del_\nu\left(\bCN\gamma^\nu\gamma^5\CN\right) +\CO(Q^3) \,.
\end{align}
This can  be also written in a non-relativistic form as
\begin{eqnarray}\label{N N eta}
\CL_{\eta'} \rightarrow
\frac{1}{2m_{\eta'}^2}
\left(\frac{N_cg_{A}^{singlet}}{2f_\pi}\right)^2
{\cal A}_6\,.
\end{eqnarray}

\subsubsection*{Four-point contact Lagrangian for isospin singlet sector}
We now summarize the non-relativistic four-point contact lagrangian
from the isospin singlet mesons  as
\begin{align}\label{4F Lagrangian isosinglet}
\CL^{(I=0)}=\CL_{\omega}+\CL_{f_1}+\CL_{\eta'}\,.
\end{align}
These are the four-point interaction with the low energy constants
$C_S$ and $C_T$ of order $Q^0$ and $C_i$'s of order $Q^2$.
By direct comparison of (\ref{4F Lagrangian isosinglet})
with the full effective Lagrangian (\ref{eq:EFT Lagrangian with As}),
the leading order constants $C_S$ and $C_T$ have the structures
\begin{eqnarray}\label{LEC Q0 singlet}
C_S^{(I=0)}&=&
\sum_{k\geq1}
\frac{1}{m_{\omega^{(k)}}^2}\left(\frac{N_cg_{V}^{(k)singlet}}{2}\right)^2\,,\nn
C_T^{(I=0)}&=&
-\sum_{k\geq1}
\frac{1}{m_{f^{(k)}}^2}\left(\frac{N_cg_{A}^{(k)singlet}}{2}\right)^2\,
\end{eqnarray}
and  the LECs of order $Q^2$ are expressed as
\begin{eqnarray}\label{LEC Q2 singlet}
-\frac{C_1^{(I=0)}}{2}&=&
\frac{1}{4 m_\CN^2}\sum_{k\geq1}\frac{1}{2m_{\omega^{(k)}}^2}\left(\frac{N_c g_{V}^{(k)singlet}}{2}\right)^2
+\sum_{k\geq1}\frac{1}{2m_{\omega^{(k)}}^4}\left(\frac{N_c g_{V}^{(k)singlet}}{2}\right)^2 \,,\nn
\frac{C_2^{(I=0)}}{8}&=&
\frac{1}{4 m_\CN^2}\sum_{k\geq1}\frac{1}{2m_{\omega^{(k)}}^2}\left(\frac{N_c g_{V}^{(k)singlet}}{2}\right)^2 \,,\nn
-\frac{C_3^{(I=0)}}{2}&=&
\frac{1}{4 m_\CN^2}\sum_{k\geq1}\frac{1}{2m_{\omega^{(k)}}^2}\left(\frac{N_c g_{V}^{(k)singlet}}{2}\right)^2
+\sum_{k\geq1}\frac{1}{2m_{f^{(k)}}^4}\left(\frac{N_c g_{A}^{(k)singlet}}{2}\right)^2 \,,\nn
-\frac{C_4^{(I=0)}}{8}&=&
-\frac{1}{4 m_\CN^2}\sum_{k\geq1}\frac{1}{2m_{f^{(k)}}^2}\left(\frac{N_c g_{A}^{(k)singlet}}{2}\right)^2 \,,\nn
-\frac{C_5^{(I=0)}}{4}&=&
\frac{3}{4 m_\CN^2}\sum_{k\geq1}\frac{1}{2m_{\omega^{(k)}}^2}\left(\frac{N_c g_{V}^{(k)singlet}}{2}\right)^2
+\frac{1}{4 m_\CN^2}\sum_{k\geq1}\frac{1}{2m_{f^{(k)}}^2}\left(\frac{N_c g_{A}^{(k)singlet}}{2}\right)^2 \,,\nn
-\frac{C_6^{(I=0)}}{2}&=&
-\frac{1}{4 m_\CN^2}\sum_{k\geq1}\frac{1}{2m_{\omega^{(k)}}^2}\left(\frac{N_c g_{V}^{(k)singlet}}{2}\right)^2
-\frac{1}{4 m_\CN^2}\sum_{k\geq1}\frac{1}{2m_{f^{(k)}}^2}\left(\frac{N_c g_{A}^{(k)singlet}}{2}\right)^2 \nn
&&\qquad+\sum_{k\geq1}\frac{1}{2m_{f^{(k)}}^4}\left(\frac{N_c g_{A}^{(k)singlet}}{2}\right)^2
+\frac{1}{2m_\eta^{'2}}\left(\frac{N_c g_{A}^{singlet}}{2f_\pi}\right)^2 \,,\nn
-\frac{C_7^{(I=0)}}{16}&=&
\frac{1}{4 m_\CN^2}\sum_{k\geq1}\frac{1}{2m_{f^{(k)}}^2}\left(\frac{N_c g_{A}^{(k)singlet}}{2}\right)^2 \,.
\end{eqnarray}

\subsection{Isospin triplet mesons}
We move on to the contribution from isospin triplet mesons to the LECs.

\subsubsection*{$\rho$ meson}
For the isospin triplet vector meson, the interaction is
\begin{align}\label{L rho CN}
-\sum_{k\geq1}\left(\frac{g_{V}^{(k)triplet}}{2}\right)\bCN \gamma^\mu \rho^{(k)a}_\mu \tau^a \CN
+\sum_{k\geq1}\left(\frac{g_{dV}^{(k)triplet}}{2}\right)\bCN \gamma^{\mu\nu}\partial_\mu\rho^{(k)a}_\nu \tau^a\CN \,.
\end{align}
Again, solving the equation of motion for $\rho^{(k)\mu a}$, we arrive at
\begin{eqnarray}\label{eq:rho}
\rho_\mu^{(k)a}
&=& -\frac{1}{m_{\rho^{(k)}}^2}\left(\frac{g_{V}^{(k)triplet}}{2}\right)\bCN\gamma_\mu \tau^a \CN
 -\frac{1}{m_{\rho^{(k)}}^2}\left(\frac{g_{dV}^{(k)triplet}}{2}\right)\del_\nu\left(\bCN{\gamma^{\nu}}_\mu \tau^a\CN\right)\nn
&&{}\, -\frac{1}{m_{\rho^{(k)}}^4}\left(\frac{g_{V}^{(k)triplet}}{2}\right)\del^2\left(\bCN\gamma_\mu \tau^a\CN\right)
\,+\CO(Q^3)\,.
\end{eqnarray}
By substitution this into (\ref{L rho CN})
we get the $\rho$ meson exchange interaction
\begin{align}\label{N N rho relativistic}
\CL_{\rho}
=&
\sum_{k\geq1}\frac{1}{2m_{\rho^{(k)}}^2}
\left(\frac{g_{V}^{(k)triplet}}{2}\right)^2\bCN\gamma_{\mu}\tau^a\CN \bCN\gamma^\mu\tau^a\CN \nn
&+\sum_{k\geq1}\frac{1}{2m_{\rho^{(k)}}^4}
\left(\frac{g_{V}^{(k)triplet}}{2}\right)^2\bCN\gamma_{\mu}\tau^a\CN \partial^2\left(\bCN\gamma^\mu\tau^a\CN\right) \nn
&+\sum_{k\geq1}\frac{1}{m_{\rho^{(k)}}^2}
\left(\frac{g_{V}^{(k)triplet}}{2}\right)\left(\frac{g_{dV}^{(k)triplet}}{2}\right)
\bCN\gamma_{\mu}\tau^a\CN \partial_\nu\left(\bCN\gamma^{\nu\mu}\tau^a\CN\right)\nn
&+\sum_{k\geq1}\frac{1}{2m_{\rho^{(k)}}^2}
\left(\frac{g_{dV}^{(k)triplet}}{2}\right)^2
\partial_\nu\left(\bCN{\gamma^\nu}_\mu\tau^a\CN\right)\partial_\lambda\left(\bCN\gamma^{\lambda\mu}\tau^a\CN\right)
\,+\CO(Q^3)\,.
\end{align}
From (\ref{eq:operators triplet}), the non-relativistic reduction of this becomes
\begin{align}\label{N N rho}
\CL_{\rho} \rightarrow &
-\sum_{k\geq1}
\frac{1}{2m_{\rho^{(k)}}^2}
\left(\frac{g_{V}^{(k)triplet}}{2}\right)^2 {\cal A}_S \nn
&+\frac{1}{4 m_\CN^2}\sum_{k\geq1}\frac{1}{2m_{\rho^{(k)}}^2}
\left(\frac{g_{V}^{(k)triplet}}{2}\right)^2
\left( -3{\cal A}_1 - 5{\cal A}_2 - {\cal A}_3 - 9{\cal A}_5 + {\cal A}_6 - {\cal A}_7 \right) \nn
&+\sum_{k\geq1}
\frac{1}{2m_{\rho^{(k)}}^4}
\left(\frac{g_{V}^{(k)triplet}}{2}\right)^2 \left( -{\cal A}_1 -2{\cal A}_2 \right) \nn
&+\frac{1}{2m_\CN}\sum_{k\geq1}\frac{1}{m_{\rho^{(k)}}^2}
\left(\frac{g_{V}^{(k)triplet}}{2}\right)\left(\frac{g_{dV}^{(k)triplet}}{2}\right)
\left( -{\cal A}_1 - 4{\cal A}_2 - {\cal A}_3 + 6{\cal A}_5 +{\cal A}_6 - {\cal A}_7 \right) \nn
&+\sum_{k\geq1}\frac{1}{2m_{\rho^{(k)}}^2}
\left(\frac{g_{dV}^{(k)triplet}}{2}\right)^2
\left( -2{\cal A}_2 -{\cal A}_3 +{\cal A}_6 - {\cal A}_7 \right)\,.
\end{align}

\subsubsection*{$a_1$ meson}
For the isospin triplet axial-vector meson $a_1$,
the interaction is
\begin{eqnarray}
-\sum_{k\geq1}
\left(\frac{g_{A}^{(k)triplet}}{2}\right)
\bCN\gamma^{\mu}\gamma^5 a_\mu^{(k)a}\tau^a\CN \,.
\end{eqnarray}
By integrating out $a_1$ meson, we have,
\begin{eqnarray}\label{N N a relativistic}
\CL_{a_1}&=&
\sum_{k\geq1}
\frac{1}{2m_{a^{(k)}}^2}
\left(\frac{g_{A}^{(k)triplet}}{2}\right)^2
\bCN\gamma^\mu\gamma^5\tau^a\CN \bCN\gamma_\mu\gamma^5\tau^a\CN \nn
&&+\sum_{k\geq1}
\frac{1}{2m_{a^{(k)}}^4}
\left(\frac{g_{A}^{(k)triplet}}{2}\right)^2
\bCN \gamma^\mu\gamma^5\tau^a\CN\del^2\left(\bCN \gamma_\mu\gamma^5 \tau^a\CN\right) \nn
&&+\sum_{k\geq1}
\frac{1}{2m_{a^{(k)}}^4}
\left(\frac{g_{A}^{(k)triplet}}{2}\right)^2
\del_\mu\left(\bCN \gamma^\mu\gamma^5 \tau^a\CN\right)\del_\nu\left(\bCN \gamma^\nu\gamma^5 \tau^a\CN\right)
\,+\CO(Q^3)\,.
\end{eqnarray}
From (\ref{eq:operators triplet}), the non-relativistic contact interaction term reads
\begin{eqnarray}\label{N N a}
\CL_{a_1}&\rightarrow&
\sum_{k\geq1}
\frac{1}{2m_{a^{(k)}}^2}
\left(\frac{g_{{\cal A}}^{(k)triplet}}{2}\right)^2 \left(3{\cal A}_S-2{\cal A}_T\right) \nn
&&+\frac{1}{4 m_\CN^2}\sum_{k\geq1}\frac{1}{2m_{a^{(k)}}^2}
\left(\frac{g_{{\cal A}}^{(k)triplet}}{2}\right)^2
\left( -{\cal A}_1 + {\cal A}_2 -{\cal A}_3 + 2{\cal A}_4 - 3{\cal A}_5 +5{\cal A}_6 - 2{\cal A}_7 \right) \nn
&&+\sum_{k\geq1}\frac{1}{2m_{a^{(k)}}^4}
\left(\frac{g_{{\cal A}}^{(k)triplet}}{2}\right)^2
\left( -4{\cal A}_2 -{\cal A}_3 -2{\cal A}_4 - {\cal A}_6 + {\cal A}_7 \right)\,.
\end{eqnarray}

\subsubsection*{Four-point contact Lagrangian for isotriplet sector}
The non-relativistic  four nucleon contact Lagrangian
from the isospin triplet meson is then
\begin{align}\label{4F Lagrangian isotriplet}
\CL^{(I=1)}=\CL_{\rho}+\CL_{a_1}\,.
\end{align}
Again, by direct comparison between (\ref{eq:EFT Lagrangian with As}) and (\ref{4F Lagrangian isotriplet})
we obtain the leading order constants $C_S$ and $C_T$
\begin{align}\label{LEC Q0 triplet}
-\frac{1}{2}C_S^{(I=1)}=&
-\sum_{k\geq1}\frac{1}{2m_{\rho^{(k)}}^2}\left(\frac{g_{V}^{(k)triplet}}{2}\right)^2
+3\sum_{k\geq1}\frac{1}{2m_{a^{(k)}}^2}\left(\frac{g_{A}^{(k)triplet}}{2}\right)^2\,,\nn
-\frac{1}{2}C_T^{(I=1)}=&
-2\sum_{k\geq1}\frac{1}{2m_{a^{(k)}}^2}\left(\frac{g_{A}^{(k)triplet}}{2}\right)^2\, .
\end{align}
The LECs of order $Q^2$ are given by
\begin{align}\label{LEC Q2 triplet}
-\frac{C_1^{(I=1)}}{2}=&
  -3\frac{1}{4 m_\CN^2}\sum_{k\geq1}\frac{1}{2m_{\rho^{(k)}}^2}\left(\frac{g_{V}^{(k)triplet}}{2}\right)^2
  -\sum_{k\geq1}\frac{1}{2m_{\rho^{(k)}}^4}\left(\frac{g_{V}^{(k)triplet}}{2}\right)^2 \nn
 &-\frac{1}{2 m_\CN}\sum_{k\geq1}\frac{1}{m_{\rho^{(k)}}^2}\left(\frac{g_{V}^{(k)triplet}}{2}\right)\left(\frac{g_{dV}^{(k)triplet}}{2}\right)
  -\frac{1}{4 m_\CN^2}\sum_{k\geq1}\frac{1}{2m_{a^{(k)}}^2}\left(\frac{g_{A}^{(k)triplet}}{2}\right)^2\,,\nn
\frac{C_2^{(I=1)}}{8}=&
  -5\frac{1}{4 m_\CN^2}\sum_{k\geq1}\frac{1}{2m_{\rho^{(k)}}^2}\left(\frac{g_{V}^{(k)triplet}}{2}\right)^2
  -2\sum_{k\geq1}\frac{1}{2m_{\rho^{(k)}}^4}\left(\frac{g_{V}^{(k)triplet}}{2}\right)^2 \nn
 &-4\frac{1}{2 m_\CN}\sum_{k\geq1}\frac{1}{m_{\rho^{(k)}}^2}\left(\frac{g_{V}^{(k)triplet}}{2}\right)\left(\frac{g_{dV}^{(k)triplet}}{2}\right)
  -2\sum_{k\geq1}\frac{1}{2m_{\rho^{(k)}}^2}\left(\frac{g_{dV}^{(k)triplet}}{2}\right)^2 \nn
 &+\frac{1}{4 m_\CN^2}\sum_{k\geq1}\frac{1}{2m_{a^{(k)}}^2}\left(\frac{g_{A}^{(k)triplet}}{2}\right)^2
  -4\sum_{k\geq1}\frac{1}{2m_{a^{(k)}}^4}\left(\frac{g_{A}^{(k)triplet}}{2}\right)^2\,,\nn
-\frac{C_3^{(I=1)}}{2}=&
  -\frac{1}{4 m_\CN^2}\sum_{k\geq1}\frac{1}{2m_{\rho^{(k)}}^2}\left(\frac{g_{V}^{(k)triplet}}{2}\right)^2
  -\frac{1}{2 m_\CN}\sum_{k\geq1}\frac{1}{m_{\rho^{(k)}}^2}\left(\frac{g_{V}^{(k)triplet}}{2}\right)\left(\frac{g_{dV}^{(k)triplet}}{2}\right) \nn
 &-\sum_{k\geq1}\frac{1}{2m_{\rho^{(k)}}^2}\left(\frac{g_{dV}^{(k)triplet}}{2}\right)^2
  -\frac{1}{4 m_\CN^2}\sum_{k\geq1}\frac{1}{2m_{a^{(k)}}^2}\left(\frac{g_{A}^{(k)triplet}}{2}\right)^2 \nn
 &-\sum_{k\geq1}\frac{1}{2m_{a^{(k)}}^4}\left(\frac{g_{A}^{(k)triplet}}{2}\right)^2 \,,\nn
-\frac{C_4^{(I=1)}}{8}=&
  2\frac{1}{4 m_\CN^2}\sum_{k\geq1}\frac{1}{2m_{a^{(k)}}^2}\left(\frac{g_{A}^{(k)triplet}}{2}\right)^2
  -2\sum_{k\geq1}\frac{1}{2m_{a^{(k)}}^4}\left(\frac{g_{A}^{(k)triplet}}{2}\right)^2 \,,\nn
-\frac{C_5^{(I=1)}}{4}=&
  -9\frac{1}{4 m_\CN^2}\sum_{k\geq1}\frac{1}{2m_{\rho^{(k)}}^2}\left(\frac{g_{V}^{(k)triplet}}{2}\right)^2
  +6\frac{1}{2 m_\CN}\sum_{k\geq1}\frac{1}{m_{\rho^{(k)}}^2}\left(\frac{g_{V}^{(k)triplet}}{2}\right)\left(\frac{g_{dV}^{(k)triplet}}{2}\right) \nn
 &-3\frac{1}{4 m_\CN^2}\sum_{k\geq1}\frac{1}{2m_{a^{(k)}}^2}\left(\frac{g_{A}^{(k)triplet}}{2}\right)^2 \,,\nn
-\frac{C_6^{(I=1)}}{2}=&
  \frac{1}{4 m_\CN^2}\sum_{k\geq1}\frac{1}{2m_{\rho^{(k)}}^2}\left(\frac{g_{V}^{(k)triplet}}{2}\right)^2
  +\frac{1}{2 m_\CN}\sum_{k\geq1}\frac{1}{m_{\rho^{(k)}}^2}\left(\frac{g_{V}^{(k)triplet}}{2}\right)\left(\frac{g_{dV}^{(k)triplet}}{2}\right) \nn
 &+\sum_{k\geq1}\frac{1}{2m_{\rho^{(k)}}^2}\left(\frac{g_{dV}^{(k)triplet}}{2}\right)^2
  +5\frac{1}{4 m_\CN^2}\sum_{k\geq1}\frac{1}{2m_{a^{(k)}}^2}\left(\frac{g_{A}^{(k)triplet}}{2}\right)^2 \nn
 &-\sum_{k\geq1}\frac{1}{2m_{a^{(k)}}^4}\left(\frac{g_{A}^{(k)triplet}}{2}\right)^2 \,,\nn
-\frac{C_7^{(I=1)}}{16}=&
  -\frac{1}{4 m_\CN^2}\sum_{k\geq1}\frac{1}{2m_{\rho^{(k)}}^2}\left(\frac{g_{V}^{(k)triplet}}{2}\right)^2
  -\frac{1}{2 m_\CN}\sum_{k\geq1}\frac{1}{m_{\rho^{(k)}}^2}\left(\frac{g_{V}^{(k)triplet}}{2}\right)\left(\frac{g_{dV}^{(k)triplet}}{2}\right) \nn
 &-\sum_{k\geq1}\frac{1}{2m_{\rho^{(k)}}^2}\left(\frac{g_{dV}^{(k)triplet}}{2}\right)^2
  -2\frac{1}{4 m_\CN^2}\sum_{k\geq1}\frac{1}{2m_{a^{(k)}}^2}\left(\frac{g_{A}^{(k)triplet}}{2}\right)^2 \nn
 &+\sum_{k\geq1}\frac{1}{2m_{a^{(k)}}^4}\left(\frac{g_{A}^{(k)triplet}}{2}\right)^2 \,.
\end{align}

\section{Low energy constants at large $N_c$}
Before we calculate the LECs numerically, we investigate the structure of them in the large $\lambda$
and large $N_c$ limit.

The leading large $N_c$ and large $\lambda$ scaling of coupling constants are:
for pseudo-scalars ($\varphi=\pi,\eta'$)
\begin{eqnarray}
\frac{g_{\pi\CN\CN}}{2m_\CN} M_{KK} &=& \frac{g_A^{triplet}}{2f_\pi}M_{KK}
\simeq \frac{2\cdot 3\cdot\pi}{\sqrt{5}} \times \sqrt{\frac{N_c}{\lambda}},
\nonumber \\
\frac{g_{\eta' \CN\CN}}{2m_\CN} M_{KK} &=& \frac{N_c g_A^{singlet}}{2f_\pi}M_{KK}
\simeq \sqrt{\frac{3^9}{2}} \pi^2 \times \frac{1}{\lambda N_c} \sqrt{\frac{N_c}{\lambda}},
\end{eqnarray}
for vector mesons ($v=\rho^{(k)},\omega^{(k)}$)
\begin{eqnarray}
g_{\rho^{(k)}\CN\CN} &=& \frac{g_V^{(k)triplet}}{2}
\simeq \sqrt{2\cdot3^3\cdot\pi^3}\:  \hat{\psi}_{(2k-1)}(0)
\times \frac{1}{N_c} \sqrt{\frac{N_c}{\lambda}} ,
\nonumber \\
g_{\omega^{(k)}\CN\CN} &=& \frac{N_c g_V^{(k)singlet}}{2}
\simeq  \sqrt{2\cdot3^3\cdot\pi^3}\: \hat{\psi}_{(2k-1)}(0) \times \sqrt{\frac{N_c}{\lambda}},
\nonumber \\
\frac{\tilde{g}_{\rho^{(k)}\CN\CN}}{2m_\CN}M_{KK} &=&
\frac{g_{dV}^{(k)triplet}M_{KK}}{2} \simeq
 \sqrt{\frac{2^2 \cdot3^2\cdot\pi^3}{5}}\:  \hat{\psi}_{(2k-1)}(0) \times
\sqrt{\frac{N_c}{\lambda}} ,
\end{eqnarray}
and for axial vector mesons ($a_1^{(k)},f_1^{(k)}$),
\begin{eqnarray}
g_{a^{(k)}\CN\CN} &\equiv& \frac{g_A^{(k)triplet}}{2}
\simeq \sqrt{\frac{2^2\cdot 3^2\cdot\pi^3}{5}}\:  \hat{\psi}_{(2k)}\,'(0)
\times \sqrt{\frac{N_c}{\lambda}} ,
\nonumber \\
g_{f^{(k)}\CN\CN} &\equiv& \frac{N_c g_A^{(k)singlet}}{2}
\simeq \sqrt{\frac{3^9 \cdot\pi^5}{2}} \: \hat{\psi}_{(2k)}\,'(0)
\times \frac{1}{\lambda N_c} \sqrt{\frac{N_c}{\lambda}} .
\end{eqnarray}
Therefore, the large $N_c$ and large $\lambda$ leading contributions
arise from the following couplings
\begin{eqnarray}\label{eq:leading classes of couplings}
\frac{g_{\pi{\cal NN}}M_{KK}}{2m_{\cal N}}
\sim
g_{\omega^{(k)}\cal NN}
\sim
\frac{\tilde g_{\rho^{(k)}\cal NN}M_{KK}}{2m_{\cal N}}
\sim
g_{a^{(k)}\cal NN}
\sim \sqrt{\frac{N_c}{\lambda}}\,.
\end{eqnarray}

\subsection{Large $N_c$ leading order with no derivative}
In this limit, we consider the leading order of large $N_c$ and
the leading  $Q^2$ order for the  four-point interactions.
As in (\ref{eq:leading classes of couplings}), the relevant mesons
in this limit are $\omega$, $\rho$ and $a_1$.

\subsubsection*{Isospin singlet}
In this $Q^2\rightarrow 0$ limit we ignore $\del^2$ terms
and corresponding Dirac spinor $\CN$ is given in (\ref{eq:CN N 0}).
For the isosinglet vector meson $\omega$ case, therefore by (\ref{N N omega}),
the $\omega$ meson exchange interaction becomes
\begin{align}
\CL_{\omega}
\rightarrow
-\sum_{k\geq1}\frac{1}{2m_{\omega^{(k)}}^2}\left(\frac{N_c g_{V}^{(k)singlet}}{2}\right)^2
O_S\,.
\end{align}
Then the four-point interaction from isosinglet sector takes the form
\begin{eqnarray}
\CL^{(I=0)}
=\CL_{\omega}
= -\frac{1}{2}C_S O_S -\frac{1}{2}C_T O_T\, ,
\end{eqnarray}
where
\begin{align}
C_S^{(I=0)}&=\sum_{k\geq1}
\frac{1}{m_{\omega^{(k)}}^2}\left(\frac{N_cg_{V}^{(k)singlet}}{2}\right)^2,\nn
C_T^{(I=0)}&=0.
\end{align}

\subsubsection*{Isospin triplet}

We see from (\ref{eq:leading classes of couplings})
that the dominant cubic couplings for the isotriplet vector meson $\rho$ in the large $N_c$ limit
is the tensor coupling only.
In this leading $Q^2\rightarrow 0$ limit, (\ref{N N rho}) shows that
$\rho$ meson part has no contribution. Therefore
\begin{equation}
\CL_{\rho} \simeq 0
\end{equation}
in this limit.
On the other hand, for the isotriplet axial vector meson $a_1$,
(\ref{N N a}) shows that the leading order $a_1$ meson interaction is
\begin{align}
\CL_{a_1}
\rightarrow \sum_{k\geq1}
\frac{1}{2m_{a^{(k)}}^2}\left(\frac{g_{A}^{(k)triplet}}{2}\right)^2
O_T\,.
\end{align}
Then we can write the leading $Q^0$ order
four-point interactions from the isotriplet mesons sector
in the large $N_c$ and large $\lambda$ limit as
\begin{eqnarray}
\CL^{(I=1)}
=\CL_{\rho}+\CL_{a_1}
= -\frac{1}{2}C_S O_S -\frac{1}{2}C_T O_T
\end{eqnarray}
where
\begin{eqnarray}
C_S^{(I=1)}&=&0,\nn
C_T^{(I=1)}&=&-\sum_{k\geq1}
\frac{1}{m_{a^{(k)}}^2}\left(\frac{g_{A}^{(k)triplet}}{2}\right)^2.
\end{eqnarray}

\subsection{Large and finite $N_c$ up to $Q^2$ order}
For the order of $Q^2$ we include both
$\CL^{(0)}$ and $\CL^{(2)}$ with the low energy constants (LECs) $C_i$'s.
In the large $N_c$ limit, the nucleon mass $m_\CN$ is proportional to
$N_c$ and therefore terms with $1/m_\CN^2$
or $1/m_\CN$ in (\ref{LEC Q2 singlet}) and (\ref{LEC Q2 triplet})
will be suppressed relative to ones without such  terms.

\subsubsection*{Isospin singlet}
For the large $N_c$ limit, from (\ref{eq:leading classes of couplings}) we need only
\begin{eqnarray}
\CL^{(I=0)}=\CL_{\omega}
\end{eqnarray}
where $\CL_{\omega}$ does not include the leading $1/{4m_\CN^2}$ term
in (\ref{N N eta}).
For a finite $N_c$, we have
\begin{eqnarray}
\CL^{(I=0)}=\CL_{\omega}+\CL_{f_1}+\CL_{\eta'}\,,
\end{eqnarray}
which are given in (\ref{N N omega}), (\ref{N N f}) and (\ref{N N eta}), respectively.
The corresponding LECs are (\ref{LEC Q0 singlet}) and (\ref{LEC Q2 singlet}).

\subsubsection*{Isospin triplet}
For the large $N_c$ limit, from (\ref{eq:leading classes of couplings}) we need only
\begin{eqnarray}
\CL^{(I=1)}=\CL_{\partial\rho}+\CL_{a_1}\,,
\end{eqnarray}
where $\CL_{\partial\rho}$ refers to  only the term with $(g_{dV}^{(k)triplet})^2$ in (\ref{N N rho}),
and  $\CL_{a\CN\CN}$ does not include the leading $1/{4m_\CN^2}$ term from (\ref{N N a}).
For a finite $N_c$, we have
\begin{eqnarray}
\CL^{(I=1)}=\CL_{\rho}+\CL_{a_1}\,,
\end{eqnarray}
which are shown in (\ref{N N rho}) and (\ref{N N a}), respectively
and the corresponding LECs are given by (\ref{LEC Q0 triplet}) and (\ref{LEC Q2 triplet}).

\section{Numerical results}
Now we are at the stage of the numerical evaluation of
the LECs for a finite $N_c$ and $\lambda$. For this we need to calculate the masses and coupling constants in
the 4D meson-baryon Lagrangian.
For illustration purpose, we will take $\lambda N_c=50$ as in \cite{Hong:2007kx, Hong:2007ay, Kim:2009sr}.

\subsection{Gauge and baryonic profile functions}

For efficient numerical estimates, we introduce dimensionless variables
$\tilde w=M_{KK}w$, $\tilde U={U/U_{KK}}$, and $\tilde z={z/ U_{KK}}$.
These are related as
\begin{equation}
\tilde w = \int_0^{\tilde z}\, \frac{d\tilde z}{\left[1+\tilde z^2\right]^{2/3}}
= \frac{3}{2}\int_1^{\tilde U}\,\frac{d\tilde U}{\sqrt{\tilde U^3 -1}}\:.
\end{equation}
In terms of these variables,
we solve the eigenvalue equations \cite{sakai-sugimoto}
\begin{equation}\label{eq:psi eigenfunc}
-(1+\tilde{z}^2)^{1/3}\partial_{\tilde{z}}((1+\tilde{z}^2)\partial_{\tilde{z}}\psi_{(n)})=\lambda_n\psi_{(n)}
\end{equation}
for the gauge sector profile functions $\psi_{(n)}$.
To compute (\ref{eq:An Bn Cn}),
we introduce re-scaled gauge profile function $\hat \psi_{(n)}(\tilde w)$
that is defined as \cite{Kim:2009sr}
\begin{equation}
\hat \psi_{(n)}(\tilde w)=\sqrt{\frac{216\pi^3}{\lambda N_c}}\psi_{(n)}(w)\,.
\end{equation}
In addition,
the conventional choice of the zero mode gauge field profile function
can be $\partial_{\tilde{z}}\psi_{(0)}(0)=1/\pi$
and we define
\begin{equation}\label{psi_0 scaled}
\hat \psi_{(0)}(\tilde{z})= \frac{1}{\pi}\tan^{-1}\tilde{z} \,.
\end{equation}
For the baryonic sector, we have
\begin{equation}
m_\CB(z)=m_\CN^{(0)}\cdot\tilde U+m_0^e=M_{KK}\cdot\left(\frac{\lambda N_c}{27\pi}\tilde U(\tilde
w)+\epsilon  N_c\right)
\end{equation}
with $\epsilon\equiv \sqrt{2/15}\simeq 0.37$.
After dividing (\ref{eigeneq}) by $M_{KK}^2$, we reach the eigenvalue equation for baryonic profile functions
\begin{equation}\label{redeigeneq}
\left[-\partial^2_{\tilde w} -\frac{\lambda N_c}{27\pi}\,
\partial_{\tilde w}\tilde U({\tilde w})+\left( \frac{\lambda N_c}{27\pi}\,\tilde U(\tilde w)+\epsilon N_c
\right)^2\right] f_+(\tilde w)
=\left(\frac{m_\CN}{M_{KK}}\right)^2
f_+(\tilde w)\:\, .
\end{equation}
Then by solving (\ref{eq:psi eigenfunc}) and (\ref{redeigeneq}), we can compute
(\ref{eq:An Bn Cn}) and finally get the value of LECs.

\subsection{Tower of couplings}

The baryon wave function (\ref{fiveB}) is effectively localized at $w\simeq 0$
in the large $N_c$ and large $\lambda$ limit.
The coefficient function for the derivative tensor coupling
$\bar\CB \gamma\cdot F \CB$ has the central value near $w \simeq 0$ as
\begin{align}
g_5(w)\frac{\rho_{baryon}^2}{ e^2(w)}
\simeq 0.18 \frac{N_c}{M_{KK}}.
\end{align}
Then the three numbers $A_n$, $B_n$ and $C_n$ in (\ref{eq:An Bn Cn}) become
\begin{align}
A_n &= \int_{-\infty}^{\infty} \frac{d\tilde{z}}{(1+\tilde{z}^2)^{2/3}}
 \,\left|f_+(\tilde{z})\right|^2 \psi_{(n)}(\tilde{z}) \:, \nn
B_n &= \frac{1}{M_{KK}}\cdot 0.18\, N_c \int_{-\infty}^{\infty} \frac{d\tilde{z}}{(1+\tilde{z}^2)^{2/3}}
 \, f_-^*(\tilde{z})f_+(\tilde{z}) \psi_{(n)}(\tilde{z}) \:, \nn
C_n &= 0.18\, N_c\int_{-\infty}^{\infty} d\tilde{z}
 \, \left|f_+(\tilde{z})\right|^2 \partial_{\tilde{z}}\psi_{(n)}(\tilde{z})
\end{align}
where $d{\tilde w} = d\tilde{z} / (1+\tilde{z}^2)^{2/3}$.
From (\ref{psi_0 scaled}) we also have $g_A^{singlet}=2A_0=0.137$.

In section \ref{sec:Mesons}, we have mentioned that $(g_{YM}^2N_c)N_c \sim 50$ is required
for $M_{KK}\sim m_{\CN}\sim 0.94 \mathrm{GeV}$, and for the case of
$N_c=3$, this gives $\lambda=g_{YM}^2N_c\simeq 17$.
Now if we take $\lambda N_c = 50$ with $N_c=3$,
we obtain the numerical eigenvalue $m_\CN/M_{KK}=1.91$ from (\ref{redeigeneq}).
(For other choices of $\lambda N_c$ and the corresponding eigenvalues, see \cite{Hong:2007ay}.)
But to obtain LECs, we will take more realistic $m_\CN\simeq M_{KK}$,
since the ratio $m_\CN/M_{KK}$ is meaningful just as an eigenvalue of each $\lambda N_c$ input.
The resulting $A_n$, $B_n$, $C_n$ and the coupling constants are shown in Table~\ref{table1}.

The KK towers of meson masses are listed in Table~\ref{table meson mass} from the previous work \cite{Kim:2009sr}
and the mass of $\eta'$ is given by (\ref{eta mass}).
In addition,
\begin{equation}
f_\pi=\sqrt{\frac{\lambda N_c}{54\pi^4}}M_{KK} = 0.0975 \;M_{KK}\,
\end{equation}
from \cite{sakai-sugimoto}.

\begin{table}[ht]
\begin{tabular*}{\linewidth}{@{\extracolsep{\fill}}|c|c|c|c|c|c|@{}c@{}|c|@{}c@{}|}
\hline
{\footnotesize$k$} &
{\footnotesize$A_{2k-1}\quad$} &
{\footnotesize$B_{2k-1}$} &
{\footnotesize$C_{2k-1}$} &
{\footnotesize $A_{2k}\quad$} &
{\footnotesize $C_{2k}$} &
{\footnotesize $A_{2k-1}\!+\!2C_{2k-1}$} &
{\footnotesize$2B_{2k-1}\quad$} &
{\footnotesize$2C_{2k}\!+\!A_{2k}$}
\\
 & {\tiny$=\!\!g_V^{(k)singlet}$} &  &  & {\tiny$=\!\!g_A^{(k)singlet}$}& &
{\tiny$=\!\!g_V^{(k)triplet}$} &
{\tiny$=\!\!g_{dV}^{(k)triplet}$} &
{\tiny$=\!\!g_A^{(k)triplet}$} \\
\hline
\hline
1 & 5.93 & 3.05 & -0.409 & 1.12 & 2.35 & 5.11 & 6.10 & 5.81 \\
\hline
2 & -3.21 & -1.75 & 1.01 & -1.03 & -1.98 & -1.19 & -3.49 & -4.98 \\
\hline
3 & 1.25 & 0.756 & -0.967 & 0.521 & 0.863 & -0.685 &  1.51 & 2.25 \\
\hline
4 & -0.305 & -0.222 & 0.481 & -0.149 & -0.188 & 0.658 & -0.443 & -0.526 \\
\hline
5 & 0.0401 & 0.0392 & -0.116 & 0.0206 & 0.0160 & -0.191 & 0.0783 & 0.0526 \\
\hline
\end{tabular*}
\caption{\small The values of the couplings. $A_n$, $C_n$ are dimensionless and $B_n$ are of dimension $M_{KK}^{-1}$.}
\label{table1}
\end{table}

\begin{table}[ht]
\begin{tabular}{|c|c|c|}
\hline
$\quad k\quad $ & $m_{\omega^{(k)}}=m_{\rho^{(k)}}$& $\quad m_{a^{(k)}}=m_{f^{(k)}}\quad $ \\
\hline
\hline
1 & 0.818 & 1.25 \\
\hline
2 & 1.69  & 2.13 \\
\hline
3 & 2.57  & 3.00 \\
\hline
4 & 3.44  & 3.87 \\
\hline
5 & 4.30 & 4.73 \\
\hline
\end{tabular}
\caption{\small
The tower of meson masses in unit of $M_{KK}$. }
\label{table meson mass}
\end{table}

\subsection{Low energy constants}

Now we obtain the values of the LECs at finite $N_c$
from the first five towers of couplings (Table~\ref{table1}) and the corresponding meson masses (Table~\ref{table meson mass}).
The leading order ($Q^0$) LECs, $C_S$ and $C_T$ of isosinglet and isotriplet channels are given in
(\ref{LEC Q0 singlet}) and (\ref{LEC Q0 triplet}), respectively.
In the next leading order ($Q^2$),
the LECs of isosinglet and isotriplet channels are listed in
(\ref{LEC Q2 singlet}) and (\ref{LEC Q2 triplet}).
The resulting values of the LECs are listed in Table~\ref{tb:LECs},
where we sum over the first five towers ($k=1, \ldots , 5$).\footnote{In Ref.~\cite{Hong:2007kx},
a subleading correction that shifts $N_c \rightarrow N_c + 2$ for certain types
of cubic meson-nucleon couplings was considered and found
to produce phenomenologically viable axial couplings to pion as well as  the anomalous
magnetic moment of nucleons. In the past, this shift has been shown to exist for
operators associated with the Hedgehog configuration of Skyrmions \cite{Dashen},
which in our approach manifest itself in $\bar \CB\gamma\cdot F\CB$ coupling in (\ref{fiveB}).
In Appendix~\ref{sec:Nc shifting}, we list modified results for the LECs, assuming such a shift.}

\begin{table}[ht]
\begin{tabular*}{\linewidth}{@{\extracolsep{\fill}}cccc}
{} & ($I=0$) & \textit{From} ($I=1$) & \textit{Total}\\
\hline
$C_S$ ($10^{-4}\,\mathrm{MeV}^{-2}$) & 1.44 & -0.123 & 1.32\\
$C_T$ ($10^{-4}\,\mathrm{MeV}^{-2}$) & -0.0272 & 0.156 & 0.129\\
\hline
$C_1$ ($10^{-9}\,\mathrm{MeV}^{-4}$) & -0.270 & 0.0454 & -0.225\\
$C_2$ ($10^{-9}\,\mathrm{MeV}^{-4}$) & 0.159 & -0.678 & -0.379\\
$C_3$ ($10^{-9}\,\mathrm{MeV}^{-4}$) & -0.0414 & 0.0446 & 0.0032\\
$C_4$ ($10^{-9}\,\mathrm{MeV}^{-4}$) & 0.00302 & 0.0213 & 0.0243\\
$C_5$ ($10^{-9}\,\mathrm{MeV}^{-4}$) & -0.240 & -0.113 & -0.353\\
$C_6$ ($10^{-9}\,\mathrm{MeV}^{-4}$) & 0.0311 & -0.0436 & -0.0125\\
$C_7$ ($10^{-9}\,\mathrm{MeV}^{-4}$) & -0.00603 & 0.297  & 0.291\\
\hline
\end{tabular*}
\caption{\small The low energy constants from isospin singlet ($I=0$) and Fierz-transformed isospin triplet ($I=1$) sectors.}
\label{tb:LECs}
\end{table}

\section{Summary}
We evaluated the low energy constant of four-nucleon contact interactions in an effective chiral Lagrangian in
the framework of holographic QCD. These contact interactions are essential to describe the short-range nuclear force
and also crucial to understand the bulk nuclear matter properties in the chiral Lagrangian.
We considered the Sakai-Sugimoto model with the bulk baryon field  to obtain meson-nucleon interactions,
and then we integrated out massive mesons to obtain the four-nucleon interactions in 4D.
We obtained two independent contact terms at the leading order ($Q^0$) and seven of them at the next leading order ($Q^2$),
which is consistent with the  effective chiral Lagrangian. We calculated the values of the LECs with the first five Kaluza-Klein
resonances.

It will be interesting to study some phenomenological consequences of the LECs determined in this work.
An immediate example might be to study observables in the nucleon-nucleon scattering such as phase shifts.

\acknowledgments
Y.K. thanks Hyun-Chul Kim and Matthias Schindler for useful discussion.
Y.K. and D.Y. acknowledge the Max Planck Society(MPG), the Korea Ministry of Education, Science and
Technology(MEST), Gyeongsangbuk-Do and Pohang City for the support of the Independent Junior
Research Group at APCTP.
P.Y. is supported  by the National Research Foundation of Korea
(NRF) funded by the Ministry of Education, Science and Technology
with grant number 2010-0013526, and also via
the Center for Quantum Spacetime (grant number 2005-0049409).

\appendix
\section{Non-relativistic Dirac spinor with relativistic correction}\label{sec:NonrelSpinor}
The free Dirac Lagrangian in our case is
\begin{equation}
\CL=\bar{\CN}(-i\gamma^\mu\del_\mu-im_{\CN})\CN
\label{Lfree}
\end{equation}
with the Weyl basis defined in~\cite{Hong:2007ay}. To develop the expressions
of some Dirac spinors with upper and lower parts, we can apply a
unitary transformation $\CN \rightarrow \CU\CN $ and $\gamma^\mu
\rightarrow \CU\gamma^\mu\CU^\dagger$ then the Lagrangian is
invariant under the transformation. If we adopt a unitary
transformation
\begin{equation}
\CU=\frac{1}{\sqrt{2}}\left(\begin{array}{rr} 1 & -i \\ i & -1\end{array}\right),
\nonumber
\end{equation}
then the transformed Dirac basis corresponding to (\ref{Lfree}) is
\begin{equation}\label{basis Dirac}
\gamma^0=\left(\begin{array}{rr} -i & 0 \\ 0 & i\end{array}\right),\quad
\gamma^i=\left(\begin{array}{cc} 0 & -\sigma_i \\ -\sigma_i & 0\end{array}\right),\quad
\gamma^5=\left(\begin{array}{rr} 0 & -i \\ i & 0\end{array}\right).
\end{equation}
Here we manipulate the decoupled two-component Dirac spinor field
that is, for positive energies, the upper components are large and the lower are small.
We use the Dirac-Pauli representation of the gamma matrices derived above and
follow the non-relativistic reduction manipulation in~\cite{Berestetsky:1982aq}.
Dirac's equation for a free electron is
\begin{equation}
i\frac{\del}{\del t}\CN=\left(\bmalpha\cdot\bmp+im_{\CN}^{}\beta\right)\CN
\end{equation}
where
\begin{equation}
\alpha^i=\left(\begin{array}{cc} 0 & -i\sigma^i \\ i\sigma^i & 0 \end{array}\right),\quad
\beta=\left(\begin{array}{cc} -i & 0 \\ 0 & i \end{array}\right).
\end{equation}
The the particle has its rest energy $m_{\CN}^{}$ and this should be excluded in the non-relativistic approximation,
and we denote $\CN$ by a function $\CN'$ with
\begin{equation}\label{eq:CN'}
\CN = e^{-im_{\CN}t}\CN'.
\end{equation}
Then we have
\begin{equation}
\left(i\frac{\del}{\del t}+m_{\CN}^{}\right)\CN'=\left(\bmalpha\cdot\bmp+im_{\CN}^{}\beta\right)\CN'.
\end{equation}
We can write
\begin{equation}
\CN'=\left(\begin{array}{c} N \\ h \end{array}\right).
\end{equation}
In the non-relativistic limit ($v\rightarrow 0$) the two components $h$ vanish and
this leads to an approximate equation involving $N$ only.
Then by substituting, we obtain the equations
\begin{eqnarray}\label{eq:phi}
i\frac{\del}{\del t}N &=& -i\bmsigma\cdot\bmp h, \\\label{eq:chi}
i\frac{\del}{\del t}h+2m_{\CN}^{} h &=& i\bmsigma\cdot\bmp N.
\end{eqnarray}
In the first approximation of $h$, the term $2m_{\CN}^{} h$ is dominant on the left hand side of (\ref{eq:chi}),
and this gives
\begin{equation}
h=i\frac{\bmsigma\cdot\bmp}{2m_{\CN}^{}}N=\frac{\bmsigma\cdot\nabla}{2m_{\CN}^{}}h\:,
\end{equation}
or
\begin{equation}
\left(\begin{array}{c} 0 \\ h \end{array}\right)
=-\frac{1}{2m_{\CN}^{}}\gamma^i\del_i
\left(\begin{array}{c} N \\ 0 \end{array}\right).
\end{equation}
Substitution of this into (\ref{eq:phi}) then gives the Pauli's equation.
Let us now drive the second approximation for $N$. The density is
\begin{equation}
\rho=\Nd N+h^\dag h=\Nd N
+\frac{1}{4m_{\CN}^2}(\nabla N^\dag\cdot\bmsigma)(\bmsigma\cdot\nabla N).
\end{equation}
To get the wave equation corresponding to the Sch\"{o}dinger equation,
one must assign $N$ by another two-component field $N_\NR$
such that
\begin{equation}
\int d^3x N_\NR^\dag N_\NR =
\int d^3x \left\{ \Nd N +\frac{1}{4m_{\CN}^2}(\nabla \Nd\cdot\bmsigma)(\bmsigma\cdot\nabla N) \right\}
\end{equation}
is the time independent integral. By the integration by parts, this can be rewritten as
\begin{equation}
\int d^3x N_\NR^\dag N_\NR =
\int d^3x \left\{ \Nd N -\frac{1}{8m_{\CN}^2}\left(\Nd\nabla^2N+(\nabla^2\Nd)N\right) \right\}.
\end{equation}
Then we can assign
\begin{equation*}
N_\NR=\left(1-\frac{\nabla^2}{8m_{\CN}^2}\right)N
\end{equation*}
and therefore, we finally get
\begin{equation}
N=\left(1+\frac{\nabla^2}{8m_{\CN}^2}\right)N_\NR.
\end{equation}
Collecting all together, we obtain the non-relativistic Dirac spinor with relativistic corrections,
\begin{equation}\label{eq:CN Napx}
  \CN(x) = \left( \begin{array}{c}
  1+\frac{1}{8m_{\CN}^2}\nabla^2 \\
\frac{1}{2m_{\CN}}\bmsigma\cdot\nabla
 \end{array} \right)N(x) + {\cal O}(Q^3).
\end{equation}

\section{Fierz identities}\label{sec:Fierz}
The Fierz identity for the Pauli matrices has the form
\begin{align}
\sigma^i_{\alpha\beta}\sigma^i_{\gamma\delta}
&= 2\delta_{\alpha\delta}\delta_{\gamma\beta}-\delta_{\alpha\beta}\delta_{\gamma\delta} \,.
\end{align}
From this, we can get the following relations
\begin{align}
\sigma^i_{\alpha\beta}\sigma^i_{\gamma\delta}
&= \frac{1}{2}\left( 3\delta_{\alpha\delta}\delta_{\gamma\beta}-\sigma^i_{\alpha\delta}\sigma^i_{\gamma\beta} \right) \,,\\
\delta_{\alpha\beta}\sigma^i_{\gamma\delta}
&= \frac{1}{2}\left(
\delta_{\alpha\delta}\sigma^i_{\gamma\beta}+\sigma^i_{\alpha\delta}\delta_{\gamma\beta}
+i\epsilon_{ilm}\sigma^{l}_{\alpha\delta}\sigma^{m}_{\gamma\beta}
\right) \,,\\
\sigma^i_{\alpha\beta}\delta_{\gamma\delta}
&= \frac{1}{2}\left(
\delta_{\alpha\delta}\sigma^i_{\gamma\beta}+\sigma^i_{\alpha\delta}\delta_{\gamma\beta}
-i\epsilon_{ilm}\sigma^{l}_{\alpha\delta}\sigma^{m}_{\gamma\beta}
\right) \,,\\
\sigma^i_{\alpha\beta}\sigma^j_{\gamma\delta}
&= \frac{1}{2}\,\Big(
\sigma^i_{\alpha\delta}\sigma^j_{\gamma\beta}
+\sigma^j_{\alpha\delta}\sigma^i_{\gamma\beta}
+\delta^{ij}\delta_{\alpha\delta}\delta_{\gamma\beta}
-\delta^{ij}\sigma^l_{\alpha\delta}\sigma^l_{\gamma\beta}\nn
& \qquad \qquad \qquad \qquad
+i\epsilon_{ijl}\sigma^{l}_{\alpha\delta}\delta_{\gamma\beta}
-i\epsilon_{ijl}\delta_{\alpha\delta}\sigma^{l}_{\gamma\beta}
\Big)\,.
\end{align}

\section{Axial coupling and an $\CO(1)$ correction}\label{sec:Nc shifting}

By using the equivalence of the constituent quark model (CQM) and
the Skyrmion in the large $N_c$ limit \cite{Dashen}, it has
been argued \cite{Hong:2007kx,Hong:2007ay}
that there is a universal sub-leading corrections to
various axial couplings and magnetic couplings of nucleons. This
enters some operators that originate from  $\bar\CB \gamma\cdot F
\CB$ in (\ref{fiveB}). The correction involves an additive shift
of type $N_c\rightarrow N_c+2$, and proved to be important if
one wishes to obtain phenomenologically viable pion axial couplings
and anomalous magnetic moments, for example. Although it is not
clear whether the D4-D8 holographic model, with its quenched nature,
captures this automatically, it does makes to think about such
a shift when we consider comparing numbers against data,

We anticipate the same shift of coefficient would affect other
operators descended from the same $\bar\CB \gamma\cdot F
\CB$, which will affect all leading couplings to axial vector
mesons and all tensor couplings to vector mesons. Practically
this can be achieved by multiplying the wavefunction
overlap coefficients, $B$'s and $C$'s, by a factor $(N_c + 2)/N_c= 5/3$
as
\begin{eqnarray}
g_V^{(k)triplet}&=& A_{2k-1}+2\left(\frac{5}{3}\right)C_{2k-1} \ , \nn
g_A^{(k)triplet}&=& 2\left(\frac{5}{3}\right)C_{2k}+A_{2k}\ ,\nn
g_{dV}^{(k)triplet}&=& 2\left(\frac{5}{3}\right)B_{2k-1}\ ,\nn
g_A^{triplet}&=& 4\left(\frac{5}{3}\right)C_0+2A_0
\end{eqnarray}
and other couplings $g_V^{(k)singlet}$, $g_A^{(k)singlet}$ and $g_A^{singlet}$
remain unchanged. Here we record the resulting changes in the mesons-nucleon
cubic couplings in Table~\ref{table1 shifted} and
the resulting values of the LECs in Table~\ref{tb:LECs Nc shifted}.

\begin{table}[ht]
\begin{tabular*}{\linewidth}{@{\extracolsep{\fill}}|c|c|c|c|c|c|@{}c@{}|c|@{}c@{}|}
\hline
{\footnotesize$k$} &
{\footnotesize$A_{2k-1}\quad$} &
{\footnotesize$B_{2k-1}$} &
{\footnotesize$C_{2k-1}$} &
{\footnotesize $A_{2k}\quad$} &
{\footnotesize $C_{2k}$} &
{\tiny $A_{2k-1}\!+\!2(\frac{5}{3})C_{2k-1}$} &
{\tiny $2(\frac{5}{3})B_{2k-1}\quad$} &
{\tiny $2(\frac{5}{3})C_{2k}\!+\!A_{2k}$}
\\
 & {\tiny$=\!\!g_V^{(k)singlet}$} &  &  & {\tiny$=\!\!g_A^{(k)singlet}$}& &
{\tiny$=\!\!g_V^{(k)triplet}$} &
{\tiny$=\!\!g_{dV}^{(k)triplet}$} &
{\tiny$=\!\!g_A^{(k)triplet}$} \\
\hline
\hline
1 & 5.93 & 3.05 & -0.409 & 1.12 & 2.35 & 4.57 & 10.2 & 8.94 \\
\hline
2 & -3.21 & -1.75 & 1.01 & -1.03 & -1.98 & 0.155 & -5.82 & -7.62 \\
\hline
3 & 1.25 & 0.756 & -0.967 & 0.521 & 0.863 & -1.98 &  2.52 & 3.40 \\
\hline
4 & -0.305 & -0.222 & 0.481 & -0.149 & -0.188 & 1.30 & -0.739 & -0.777 \\
\hline
5 & 0.0401 & 0.0392 & -0.116 & 0.0206 & 0.0160 & -0.345 & 0.131 & 0.0739 \\
\hline
\end{tabular*}
\caption{\small $N_c$ shifted couplings. $A_n$, $C_n$ are dimensionless and $B_n$ are of dimension $M_{KK}^{-1}$.}
\label{table1 shifted}
\end{table}

\begin{table}[ht]
\begin{tabular*}{\linewidth}{@{\extracolsep{\fill}}cccc}
{} & ($I=0$) & \textit{From} ($I=1$) & \textit{Total}\\
\hline
$C_S$ ($10^{-4}\,\mathrm{MeV}^{-2}$) & 1.44 & -0.464 & 0.976\\
$C_T$ ($10^{-4}\,\mathrm{MeV}^{-2}$) & -0.0272 & 0.369 & 0.342\\
\hline
$C_1$ ($10^{-9}\,\mathrm{MeV}^{-4}$) & -0.270 & 0.0493 & -0.221\\
$C_2$ ($10^{-9}\,\mathrm{MeV}^{-4}$) & 0.159 & -1.11 & -0.951\\
$C_3$ ($10^{-9}\,\mathrm{MeV}^{-4}$) & -0.0414 & 0.0947 & 0.0533\\
$C_4$ ($10^{-9}\,\mathrm{MeV}^{-4}$) & 0.00302 & 0.0505 & -0.0475\\
$C_5$ ($10^{-9}\,\mathrm{MeV}^{-4}$) & -0.240 & -0.185 & -0.425\\
$C_6$ ($10^{-9}\,\mathrm{MeV}^{-4}$) & 0.0311 & -0.0923 & -0.0612\\
$C_7$ ($10^{-9}\,\mathrm{MeV}^{-4}$) & -0.00603 & 0.616  & 0.610\\
\hline
\end{tabular*}
\caption{\small The low energy constants of $N_c$ shifted results.}
\label{tb:LECs Nc shifted}
\end{table}

\end{document}